\documentclass[review,12pt]{elsarticle}
\usepackage[utf8]{inputenc}
\usepackage[toc,page]{appendix}
\usepackage{amssymb}            
\usepackage{graphicx}
\usepackage{booktabs}
\usepackage{xcolor}
\usepackage{gensymb}
\usepackage{amsmath,amsfonts,amsthm,bm} 
\usepackage{mathrsfs}
\usepackage{gensymb}
\usepackage{multirow}
\usepackage[skip=6pt plus1pt, indent=40pt]{parskip}
\usepackage[margin=2cm]{geometry}
\usepackage{setspace}
\usepackage{chngcntr}
\usepackage{upgreek}
\usepackage{enumitem}
\usepackage[english]{babel}
\usepackage{microtype}
\usepackage{lineno}
\setlist[itemize]{align=parleft,left=0pt..0em}

\journal{Corrosion Science}
\usepackage{graphicx}
\graphicspath{{Figures/}}
\usepackage{caption}
\usepackage{subcaption}
\usepackage{float}
\usepackage{hyperref}
\hypersetup{colorlinks=true, linkcolor=blue, citecolor=black, filecolor=magenta, urlcolor=black}
\usepackage{soul}
\usepackage{accents}
\usepackage{color,soul}
\usepackage{color}
\usepackage{xcolor,soul}
\biboptions{sort&compress}

\begin{document}

\begin{frontmatter}

\title{A phase-field model for microbiologically influenced corrosion}

\author{Sasa Kovacevic}

\author{Emilio Mart\'{\i}nez-Pa\~neda\corref{cor1}}
\ead{emilio.martinez-paneda@eng.ox.ac.uk}

\cortext[cor1]{Corresponding author.}

\address[Ox]{Department of Engineering Science, University of Oxford, Oxford OX1 3PJ, UK}

\begin{abstract}
A phase-field-based reaction-diffusion corrosion model is developed to predict microbially influenced corrosion (MIC) in metal alloys, with a focus on anaerobic conditions and sulfate-reducing bacteria (SRB). The formulation couples microbial sulfate reduction, sulfate transport, electrochemical kinetics, material dissolution, and mechanical effects within a unified framework. Microbial activity is modelled using a Monod-type expression for sulfate consumption, whereas the mechano-chemical coupling is incorporated through an enhanced mobility relationship that captures the influence of mechanical fields on corrosion kinetics. The model is calibrated against experimental data and shows strong agreement in predicting pitting kinetics under SRB activity. Sensitivity analyses quantify the competing roles of microbial kinetics, transport, and thermodynamic driving forces in governing corrosion behaviour. The capability of the formulation to capture both MIC-induced pitting and stress-assisted corrosion across multiple length scales is demonstrated through case studies that include microstructure-sensitive simulations and structural-scale coupling with a cathodic protection (CP) model. Results show that finer grain sizes reduce pitting severity but promote faster defect propagation under mechanical loading. At the structural scale, coupling with the CP model enables predictions of defect growth under varying electrochemical conditions and over engineering-relevant length scales, as exemplified with the analysis of an offshore wind turbine monopile. CP delays pitting and suppresses crack propagation, although its effectiveness diminishes as sacrificial anodes degrade. The framework provides a predictive and computationally efficient tool for assessing MIC-induced damage over extended service times, with potential applications in the integrity and life assessment of metallic structures operating in aggressive microbial environments.\\

\end{abstract}

\end{frontmatter}

\section{Introduction} \label{sec1}

Microbiologically influenced corrosion (MIC) is one of the most severe localised forms of metal degradation, posing significant challenges across numerous engineering applications, from offshore wind turbines to wastewater utilities \cite{Xu2023}. Among the various microorganisms that could be potentially involved, sulfate-reducing bacteria (SRB) are widely recognised as the primary agents responsible for MIC \cite{GU2019}. These anaerobic bacteria form biofilms on metallic surfaces under sulfate-rich conditions, which promote their proliferation and result in highly heterogeneous electrochemical micro-environments. Through metabolic processes, SRB utilise sulfate ions (SO$_4^{2-}$) as terminal electron acceptors and produce sulfide species, which react with the metal surface and alter interfacial electrochemistry, promoting localised anodic dissolution and pitting corrosion \cite{Enning2014}.

Despite significant research on mitigation strategies, SRB-induced MIC remains a notable technological challenge, resulting in structural failures and substantial financial losses \cite{Wang2023}. The local nature of SRB-induced MIC and its interaction with the underlying microstructure and evolving mechanical fields make it a particularly challenging phenomenon to predict and prevent. Crystallographic texture (e.g., grain size and orientation) influences MIC by modulating local electrochemical heterogeneity \cite{VENEGAS2015}, leading to distinctly different MIC rates \cite{YU2023}, and establishing preferential sites for pit initiation. Moreover, the problem is worsened by the presence of mechanical loads, as SRB activity has been shown to increase the susceptibility of materials to stress-assisted corrosion and stress corrosion cracking (SCC) \cite{WU2015}. The interaction between mechanics and SRB-induced MIC is rich and complex \cite{Yan2026}, including synergistic effects that drive corrosion localisation, the formation of occluded electrochemical regions (where SRB thrive), strain-driven film rupture, and SRB-produced hydrogen and its associated embrittlement effects under local stresses. These phenomena are strongly coupled, with the interplay between microbial metabolism, mechanical loading, and microstructural heterogeneity governing behaviour at all stages in MIC. First, corrosion susceptibility becomes localised as a result of non-uniform biofilm formation due to microstructural heterogeneities \cite{WU2014a, Wu2014b}, and of local acceleration of corrosion rates due to mechanical stresses \cite{Gutman1988, CUI2021}. This leads to the formation of pits, where microbial activity is intensified, and local stresses are amplified \cite{JAVAHERDASHTI2006, Chen2022, LI2025}. These stresses accelerate corrosion kinetics, assist MIC-driven pit growth, and facilitate the pit-to-crack transition and material failure \cite{Wu2016, Wu2019, YANG2020}. There is a need to develop physically-based models that can directly simulate these processes and thus link MIC environments with the likelihood of localised corrosion failures.

While some MIC models have been proposed, these remain limited in scope \cite{MARCIALES2019}. In particular, they neglect metallurgical heterogeneity and do not capture key coupled physics, such as the arbitrary evolution of the metal-environment interface and mechano-chemical interactions \cite{Peng1994, Javaherdashti2004, AlDarbi2005, Maxwell2006, Melchers2006, Gu2009, Larsen2013, Gu2014, Haile2015, XU2016, DAWUDA2021, ANGUITA2022}. There is a need for an MIC model that can directly simulate how bacterial activity, mechanical loading and material microstructure interact, as needed to provide mechanistic predictions of localised corrosion failures. An avenue to tackle this elusive goal is provided by the phase-field method, which has recently emerged as a promising computational approach for simulating degradation processes in various materials and media \cite{martinez2024phase}, capturing complex interface dynamics across multiple length scales in uniform, pitting, and stress-assisted corrosion \cite{MAI2016, Ansari2018, Chadwick2018, LIN2021, Brewick2022, Cui2023, KOVACEVIC2023, Song2023, Makuch2024, ZENG2024, KANDEKAR2024, AMADOR2024, Lhoest2025, CHEN2026}. However, no phase-field model for microbial corrosion currently exists. Existing phase-field corrosion models have been developed primarily for purely abiotic or chemically governed systems, in which metal dissolution is driven by chemical and electrochemical reactions between the metal and its environment; i.e., they do not account for microbial metabolism and biofilm activity, features which are central to MIC. To the authors’ knowledge, this work presents the first phase-field-based reaction-diffusion model for microbially influenced corrosion that explicitly couples biological kinetics, electrochemical reactions, mass transport, and mechano-chemical interactions, enabling a unified description of MIC-induced pitting and stress-assisted corrosion.

The remainder of the paper is organised as follows. The phase-field model of microbially influenced corrosion is developed in the following section. Model calibration against experimental data under SRB conditions and the sensitivity analysis are described in Section \ref{sec3}. After calibration, the potential of this strategy to assess MIC-induced pitting and stress-assisted corrosion across multiple length scales is demonstrated through representative case studies in Section \ref{sec4}. The implications and limitations of the model, along with recommendations for future work, are discussed in Section \ref{sec5}. The manuscript ends with concluding remarks in Section \ref{sec6}.

\section{The phase-field model of microbially influenced corrosion} \label{sec2}

\subsection{Biocatalytic electrochemistry}  \label{sec21}

Microbially influenced corrosion by SRB is characterised here using the Biocatalytic Cathodic Sulfate Reduction (BCSR) theory \cite{Gu2009}. The degradation mechanism is driven by the electrochemical coupling of anodic iron oxidation and biocatalytic cathodic sulfate reduction in the cytoplasm of SRB cells
\begin{align}
\label{eqn1a}
& \text{Fe}_\mathrm{(s)} \rightarrow \text{Fe}_\mathrm{(aq)}^{2+} + 2\text{e}^- \text{ (Iron oxidation reaction)} \\
\label{eqn1b}
& \text{SO}_{4\mathrm{(aq)}}^{2-} + 9\text{H}_\mathrm{(aq)}^+ + 8\text{e}^- \rightarrow \text{HS}_\mathrm{(aq)}^- + \text{4H}_2\text{O} \text{ (BCSR reaction)}
\end{align}
The cathodic reaction is catalysed by the hydrogenase enzyme system in hydrogenase-positive SRB cells, which accelerates sulfate reduction. The iron oxidation reaction occurs outside SRB cells. In this pathway, elemental iron serves as the electron donor, whereas sulfate acts as the terminal electron acceptor during sulfate respiration \cite{GU2019}. By acquiring electrons from metal surfaces in environments lacking organic electron donors, SRB can sustain their core metabolic processes. This direct electron uptake facilitates the release of Fe$^{2+}$ ions, thereby degrading the corrosion resistance and mechanical integrity of susceptible materials. Although actual cathodic processes are more complex and involve multiple intermediate steps \cite{Xu2023}, the present model focuses on the net electrochemical effect. Moreover, extracellular electron transfer mechanisms, including those involving hydrogenases \mbox{\cite{Wang2024, Dong2026, Li2026}}, are not explicitly considered. Their net effect is incorporated through the biocatalytic cathodic sulfate reduction reaction in Eq. (\mbox{\ref{eqn1b}}).

\subsection{Kinematics and modeling assumptions}  \label{sec22}

\begin{figure}[h!]
    \centering
    \includegraphics[width = 16 cm]{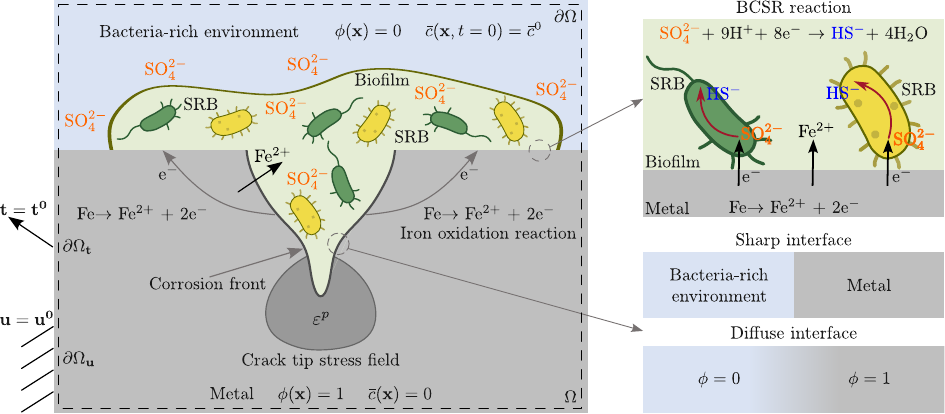}
     \captionsetup{labelfont = bf, justification = raggedright}
    \caption{Microbially influenced corrosion mechanism and diffuse interface representation of the corrosive environment (bacteria-rich environment with biofilm $\phi = 0$) and metal ($\phi =1$) phases.}
    \label{Fig1}
\end{figure}

The following independent primary variables are introduced to describe the degradation mechanism illustrated in Fig. \ref{Fig1} and the associated bioelectrochemical reactions in Eqs. (\ref{eqn1a}) and (\ref{eqn1b}). Since sulfate availability governs SRB metabolic activity, the normalised sulfate concentration $\bar{c}(\mathbf{x},t)=c/c_\mathrm{ref}$ is introduced in the model to characterise mass transport within the bulk electrolyte. Here, $c_\mathrm{ref}$ stands for the reference sulfate concentration. More details regarding nondimensionalization are given in Section \ref{sec25}. In the present formulation, sulfate is treated as the sole growth-limiting substrate \cite{Muyzer2008}, an approach frequently followed in numerical models \cite{Peng1994, AlDarbi2005}. This assumption is appropriate for anaerobic environments in which sulfate availability is the primary factor governing SRB activity and other environmental variables, such as pH, temperature, electron-donor availability, nutrient concentration, and sulfide accumulation, remain approximately constant. Consequently, the current framework does not explicitly capture changes in microbial activity arising from variations in these factors. The concentration of Fe$^{2+}$ ions is not explicitly tracked as the amount of dissolved iron can be directly inferred from sulfate reduction stoichiometry. For instance, each mole of reduced sulfate corresponds to four moles of dissolved iron. The reaction zone, interpreted as a thin SRB biofilm where reactions (\ref{eqn1a}) and (\ref{eqn1b}) take place, is assumed to be much thinner than the surrounding diffusion zone. Consequently, biofilm formation and growth are not explicitly modelled (i.e., their effect on mass transport is neglected). Sulfate consumption is assumed to occur exclusively through microbial reduction to sulfide, Eq. (\ref{eqn1b}), within this thin reaction zone and its immediate vicinity, as detailed below in Section \ref{sec24}. The depletion of SO$_{4}^{2-}$ at the interface is replenished by diffusion from the bulk electrolyte.

The evolution of the corroding interface is tracked by a continuous phase-field variable $\phi(\mathbf{x},t)$. The variable takes the value $\phi=1$ in the uncorroded metal and $\phi=0$ in the corrosive environment. The thin diffuse interface region, representing the metal-environment interface, is described with $0 < \phi < 1$. The mechanical response of the degrading metal is characterised under the assumption of small deformations (linearised kinematics) with the displacement vector $\mathbf{u}(\mathbf{x},t)$ serving as the primary kinematic variable. The overall degradation mechanism and the associated kinematic variables are summarised schematically in Fig. \ref{Fig1}.

\subsection{Thermodynamics}  \label{sec23}

The free energy functional for the heterogeneous system in Fig. \ref{Fig1} can be written as 
\begin{equation} \label{eqn5}
                   \mathscr{F} = \int_\Omega \Big[\mathcal{F}^\mathrm{chem}( \bar{c}, \phi) + \frac{1}{2}\kappa|\nabla \phi|^2 +\mathcal{F}^\mathrm{mech}(\mathbf{\nabla u},\phi) \Big]\,\text{d}\Omega,
\end{equation}
where $\mathcal{F}^\mathrm{chem}$ is the chemical free energy density, $\mathcal{F}^\mathrm{mech}$ the mechanical free energy density, $\kappa$ the gradient energy coefficient, and $\Omega$ the system domain, which includes the corroding metal and the corrosive environment consisting of sulfate ions and SRB.

The chemical free energy density $\mathcal{F}^\mathrm{chem}$ is decomposed into the chemical free energy density ascribed to the sulfate concentration and the double-well free energy density associated with the phase-field variable. Following an assumption that the interfacial region is a mixture of both phases with the same chemical diffusion potential \cite{KKS1999}, the chemical free energy density of the system can be written as (see \ref{appendixA})
\begin{equation} \label{eqn6}
\mathcal{F}^\mathrm{chem} (\bar{c}, \phi) =  \frac{1}{2} N \Psi \Big[ \bar{c} -\bar{S}_\mathrm{min} \Big(1- h(\phi) \Big) \Big]^2 + \omega g(\phi),
\end{equation}
where $N$ is the effective SRB cell density (i.e., number of SRB cells per unit volume), $\Psi$ the effective stored energy in the body of a cell, $g(\phi) = 16\phi^2(1-\phi)^2$ the double-well free energy function employed to describe the two equilibrium states for the solid ($\phi = 1$) and the liquid ($\phi = 0$) phases, $\omega$ the constant that determines the energy barrier at $\phi = 1/2$ between the two minima at $\phi = 0$ and $\phi = 1$, and $h(\phi) = \phi^3(6\phi^2-15\phi+10)$ is a monotonically increasing function that interpolates between the chemical free energies of the two phases. The expression (\ref{eqn6}) ensures that the chemical free energy density is zero in the solid phase when $\bar{c} = 0$ as SO$_{4}^{2-}$ is absent in the metal alloy. $\bar{S}_\mathrm{min} = S_\mathrm{min}/c_\mathrm{ref}$ in Eq. (\ref{eqn6}) designates the normalised minimum (limiting) sulfate concentration required for microbial activity, i.e., at or below this sulfate concentration, SRB metabolic activity and growth become negligible, leading to suppression of corrosion. The expression is defined following a Monod-type relation for describing microbial growth: $S_\mathrm{min} = K_\mathrm{m} b/(Yq-b)$ \cite{Rittmann1980}. Here, $K_\mathrm{m}$ stands for the Monod half-velocity coefficient for SRB (mol/m$^3$), $Y$ is the true yield of bacterial mass per unit of metal substrate utilised (kg/mol), $b$ denotes the specific decay or maintenance-respiration coefficient (1/s), and $q$ corresponds to the maximum specific rate of metal substrate utilisation by SRB (mol/(kg$\cdot$s)). The condition $\bar{c} > \bar{S}_\mathrm{min}$ indicates that the microbial population has a positive net growth rate, whereas the condition $\bar{c} < \bar{S}_\mathrm{min}$ yields a negative net growth rate, which eventually leads to the disappearance of the microbial population \cite{Rittmann1980}. The adopted approximation of the chemical free energy density as a parabolic function around the sulfate concentration $\bar{S}_\mathrm{min}$ admits sustained bacterial activity above the threshold concentration (see \ref{appendixA}). A sensitivity analysis is conducted in Section \ref{sec32} to examine the effects of $K_\mathrm{m}$, $Y$, $b$, $q$, $N$, and $\Psi$ on the corrosion kinetics.

The total mechanical free energy density $\mathcal{F}^\mathrm{mech}$ is expressed using linearised kinematics and following an isotropic power law hardening response within von Mises plasticity theory \cite{Simo1998} 
\begin{equation} \label{eqn12a}
                   \mathcal{F}^\mathrm{mech} (\mathbf{\nabla u},\phi) = \frac{1}{2} (\bm{\varepsilon} - \bm{\varepsilon}^{p}) : \mathbb{C} (\phi) : (\bm{\varepsilon} - \bm{\varepsilon}^{p}) + \frac{\sigma_{y} \varepsilon_y}{n+1}\Big [ \Big(1 + \frac{\varepsilon^p}{\varepsilon_y} \Big)^{n+1} - 1 \Big ],
\end{equation}
where $\bm{\varepsilon}=1/2(\nabla \mathbf{u} + (\nabla \mathbf{u})^\mathrm{T})$ is the total strain tensor, $\bm{\varepsilon}^{p}$ the plastic strain tensor, $\mathbb{C} (\phi)= h(\phi)\mathbb{C}_0$ and $\mathbb{C}_0$ are the effective and intact rank-four elastic stiffness tensors, respectively, $\sigma_{y}$ denotes the initial yield strength, $\varepsilon_y$ the initial yield strain, $n$ the hardening exponent ($0<n\le1$), and $\varepsilon^p = \sqrt{2/3} \int_{0}^{t} | \dot{\bm{\varepsilon}}^{p} |\, dt$ is the von Mises equivalent plastic strain. $h(\phi)$ acts as a degradation function that ensures that mechanical fields are confined to the solid phase and vanish smoothly within the electrolyte. The rank-four elastic stiffness tensor for the intact solid is described by isotropic linear elasticity: $\mathbb{C}_0 = \lambda\, \mathbf{I}\otimes\mathbf{I} + 2\mu\, \mathbb{I}$, where $\mathbf{I}$ is the second-order identity tensor, $\mathbb{I}$ the symmetric fourth-order identity tensor, and $\lambda$ and $\mu$ are the Lam\'e elastic constants.

\subsection{Governing equations and mechano-chemical coupling} \label{sec24}

The evolution of the corroding interface follows the usual Allen-Cahn equation for non-conserved fields \cite{ALLEN1979}
\begin{equation} \label{eqn9}
\frac{\partial \phi}{\partial t} = - L\frac{\delta \mathscr{F}}{\delta \phi} = -L\Big(\frac{\partial \mathcal{F}^{\mathrm{chem}}}{\partial\phi} - \kappa\nabla^2\phi \Big)  \quad \text{in}\quad\Omega,
\end{equation}
where $L>0$ is the interfacial mobility that controls the motion of the metal-environment interface. On the boundary $\partial\Omega$: $\kappa \mathbf{n} \cdot \nabla \phi = 0$. Following Gutman’s theory of mechano-electrochemical interactions \cite{Gutman1988,CUI2022} and considering the linear relation between the interfacial mobility and corrosion current density \cite{MAI2016, Cui2023}, the phase-field mobility is multiplicatively decomposed into three contributions \cite{CUI2021}
\begin{equation} \label{eqn10}
                   L = L_0\Big(\frac{\varepsilon^{p}}{\varepsilon_{y}} + 1 \Big)\exp\Big(\frac{\sigma_\mathrm{h} V_\mathrm{m}}{RT}\Big),
\end{equation}
where $L_0$ is the interfacial mobility in the absence of mechanical fields (stress-free corrosion). The last two factors in Eq. (\ref{eqn10}) define an amplification factor that depends on the distributions of local plastic strain $\varepsilon^{p}$ and hydrostatic stress $\sigma_\mathrm{h}$ (stress-assisted corrosion). Here, $R$ is the universal gas constant, $V_\mathrm{m}$ the molar volume of the metal, and $T$ the absolute temperature. For an alternative way of incorporating the mechanical contribution to interface kinetics, the interested reader is referred to Refs. \cite{LIN2021, KOVACEVIC2025}. The mechanical fields are obtained by solving the linear momentum balance equation for quasi-static loading 
\begin{equation} \label{eqn11}
\begin{aligned}
& \nabla \cdot \bm{\sigma} = \bm{0}  \quad\mathrm{}\quad \text{in } \Omega \\
\mathbf{t} = \mathbf{n}\cdot \bm{\sigma} = \mathbf{t}^0 \quad &\text{on}\quad \partial\Omega_{\mathbf{t}} \quad \text{and} \quad \mathbf{u} = \mathbf{u}^0 \quad\mathrm{} \text{on}\quad \partial\Omega_{\mathbf{u}},
\end{aligned}
\end{equation}
complemented by standard boundary conditions with prescribed traction $\mathbf{t}^0$ and displacement $\mathbf{u}^0$ vectors on boundaries $\partial\Omega_{\mathbf{t}}$ and $\partial\Omega_{\mathbf{u}}$. In the previous expression, $\bm{\sigma} = \partial \mathcal{F}^\mathrm{mech}/\partial(\bm{\varepsilon} - \bm{\varepsilon}^{p}) = h(\phi)\mathbf{C}_0 : (\bm{\varepsilon} - \bm{\varepsilon}^{p})$ represents the effective Cauchy stress tensor.

The  transport of sulfate ions is governed by the mass balance law
\begin{equation} \label{eqn14}
\left\{
\begin{aligned}
&\frac{\partial \bar{c}}{\partial t} = - \nabla \cdot \mathbf{J} + R ;\quad\mathrm{}\quad R = - \frac{q c_b}{c_\mathrm{ref}}\frac{\bar{c}}{(\bar{K}_\mathrm{m} + \bar{c})}p(\phi)\\
& \mathbf{J} = -M \nabla \Big(\frac{\delta \mathscr{F}}{\delta \bar{c}} \Big) = -D \nabla \bar{c} - D \bar{S}_\mathrm{min} h^{\prime}(\phi)\nabla \phi\\
\end{aligned}
\right\}\
\quad\text{in }\Omega.
\end{equation}
On the boundary $\partial\Omega$: $\mathbf{n} \cdot \mathbf{J} = 0$. $\mathbf{J}$ stands for the diffusional flux, $M$ is the mobility parameter ($M = D/(\partial^2 \mathcal{F}^\mathrm{chem} / \partial \bar{c}^2)$), $D$ denotes the effective diffusion coefficient, and $h^{\prime}(\phi) = \partial h(\phi) / \partial \phi$. The effective diffusion coefficient is interpolated with the phase-field variable: $D = D^s h(\phi)+(1-h(\phi))D^l$, where $D^l$ and $D^s$ stand for the diffusion coefficients of sulfate ions in the liquid and solid phases. $D^s \ll D^l$ is enforced to suppress the transport within the metal phase. $R$ in Eq. (\ref{eqn14}) is the reaction rate that accounts for sulfate consumption by SRB cells. It is defined following the Monod relationship for the rate of sulfate utilisation \cite{Peng1994}. Here, $\bar{K}_\mathrm{m} = K_\mathrm{m}/c_\mathrm{ref}$ is the normalised Monod half-velocity coefficient for SRB, $c_b$ the bacterial density in the biofilm in units of kg/m$^3$, and $p(\phi) = 4\phi(1-\phi)$ the function introduced to limit annihilation of sulfate ions to a limited area close to the metal-electrolyte interface (i.e., the reaction zone). The reaction term influences interface evolution indirectly by affecting the sulfate concentration field, which enters the chemical free energy density.

\subsection{Parameter identification} \label{sec25p}

The present model consists of physical material properties, microbial-electrochemical (biological) parameters, and phase-field model parameters. The mechanical, material, and transport properties of the metal–electrolyte system can be directly obtained from experimental data. These include the elastic constants ($\lambda$, $\mu$), plasticity parameters ($\sigma_{y}$, $n$), molar volume $V_\mathrm{m}$, and the diffusion coefficient of sulfate ions in the environment $D^l$. The kinetics of sulfate consumption and the chemical free energy density are governed by biologically measurable quantities: $K_\mathrm{m}$, $b$, $Y$, $q$, $c_b$. These parameters depend on the SRB strain, environmental conditions (e.g., pH and temperature), and the surrounding environment (e.g., freshwater, landfill, and marine environments). They can be selected accordingly from experimental studies. The effective stored cell energy $\Psi$ defines the energetic contribution of microbial activity to the chemical free energy density, which can be determined from the calorific content \cite{Wang1976} and the weight of SRB cells. It is treated as a constant representative value for a given SRB strain in the present investigation. The phase-field model parameters $\kappa$ and $\omega$ are related to the interfacial energy $\Gamma$ (a physical quantity) and the chosen diffuse interface thickness $\ell$ through $\kappa =3\Gamma\ell/2$ and $\omega=3\Gamma/(4\ell)$ \cite{KOVACEVIC2020}.

The only fitting parameter that needs to be calibrated against experimental data is the interfacial mobility prefactor $L_0$ in Eq. (\ref{eqn10}). This equation is motivated by Gutman's mechano-chemical theory \cite{Gutman1988}, which provides a phenomenological yet physical mechano-chemical coupling in which mechanical fields enter dissolution kinetics in a physically interpretable manner. Following the linear proportionality between interface velocity and corrosion current density $i_a$ (see \ref{appendixB}), $L_0$ can be interpreted as a kinetic coefficient linking electrochemical dissolution rate to phase-field interface motion. Its order of magnitude and lower bound can be estimated from the sharp-interface analysis and subsequently refined by fitting to experimental data. This calibration process sets the absolute time scale of the problem without altering the functional dependence of interface velocity on mechanical fields. See \ref{appendixB} for more details.

\subsection{Elements of dimensional analysis} \label{sec25}

Computations are carried out in a dimensionless domain, reducing the number of independent variables and revealing the governing physical mechanisms. The characteristic length scale of the problem is given by the nominal diffuse interface thickness $\ell$, as inferred from the phase-field equation (\ref{eqn9}). The energy barrier parameter $\omega$, which depends on $\ell$, is therefore adopted as the reference energy density. Mass transport occurs only in the electrolyte at a rate governed by $D^l$. This naturally sets the characteristic diffusion time scale. Concentrations are normalised with respect to a reference sulfate concentration $c_\mathrm{ref}$. Choosing $c_\mathrm{ref} = S_\mathrm{min}$, where $S_\mathrm{min}$ is the threshold sulfate concentration required for sustained microbial activity, leads to $\bar{S}_\mathrm{min} = 1$, which simplifies the formulation.

The analysis of the phase-field and diffusion equation yields the characteristic time for interface movement, mass transport (diffusion) within the bulk electrolyte, and the reaction rate for the consumption of sulfate ions
\begin{equation} \label{eqn15}
                   \tau_\phi = \frac{1}{L\omega} \quad\mathrm{}\quad \tau_d = \frac{\ell^2}{D^l} \quad\mathrm{}\quad \tau_r = \frac{S_\mathrm{min}}{q c_b}.
\end{equation}
The reaction time scale $\tau_r$ is controlled exclusively by bacterial parameters and represents the time required for an SRB population of density $c_b$ to consume the minimum sulfate concentration required for positive microbial activity. Small values of $\tau_r$ (high rate of metal substrate utilisation $q$) indicate highly active bacteria and correspond to rapid sulfate consumption with increased electron demand. Large $\tau_r$ values (a high threshold level for sustained microbial activity) indicate sluggish microbial activity and slow sulfate reduction.

Denoting the nondimensional quantities with a bar, e.g., $\bar{t} = t/\tau_d$ and $\bar{\nabla} = \ell \nabla$, the set of the nondimensional governing equations is written as
\begin{equation} \label{eqn16}
\left\{
\begin{aligned}
& \frac{\partial \phi}{\partial \bar{t}} = -\tau \Big[\Lambda \Big(\bar{c} - 1 + h(\phi) \Big)  h^\prime(\phi) + g^\prime(\phi) - 2 \bar{\nabla}^2 \phi \Big]\\
&\frac{\partial \bar{c}}{\partial \bar{t}} = \bar{\nabla} \cdot \Big( \bar{D} \bar{\nabla} \bar{c} + \bar{D} h^{\prime}(\phi)\bar{\nabla} \phi \Big) - D_a \frac{\bar{c}}{\bar{K}_\mathrm{m} + \bar{c}}p(\phi)\\
& \bar{\nabla} \cdot \bar{\bm{\sigma}} = \bm{0}\\ 
\end{aligned}
\right\}\
\quad\text{in }\bar{\Omega}.
\end{equation}
Three dimensionless parameters characterise the problem. The ratio $\tau = \tau_d / \tau_\phi$ sets the time scale and determines which physical process governs the overall corrosion rate. For instance, for the case of $\tau > 1$, the corrosion rate is governed by the rate of the transport of sulfate ions from the environment to the SRB at or in the immediate vicinity of the corroding surface (diffusion-controlled corrosion). The opposite condition, commonly referred to as activation-controlled corrosion, is an interface reaction-controlled process, in which the dissolution rate is limited by the availability of sulfate ions (the electron acceptor) at the corroding interface. The competition between transport and microbial reaction kinetics is characterised by the Damk\"ohler number $D_a = \tau_d/\tau_r$. For $D_a \gg 1$, sulfate is consumed faster than it is supplied, leading to depletion near the interface and transport-limited behaviour. Large $D_a$ values are associated with high SRB densities and utilisation rates. For $D_a \ll 1$, sulfate is abundant and microbial activity limits the process, resulting in biokinetically controlled corrosion. In this case, corrosion proceeds at the maximum biological potential. The energy parameter $\Lambda = N\Psi /\omega$ defines the relative magnitude of the chemical driving force and sets the energy scale governing interface evolution. The code developed, together with example case studies and documentation, will be available at \url{https://mechmat.web.ox.ac.uk/codes} after article acceptance.

\section{Model calibration and sensitivity analysis} \label{sec3}

\subsection{Model calibration} \label{sec31}

\subsubsection{Experimental and computational setup} \label{sec311}

\begin{figure}[h!]
    \centering
    \includegraphics[width = 8.5 cm]{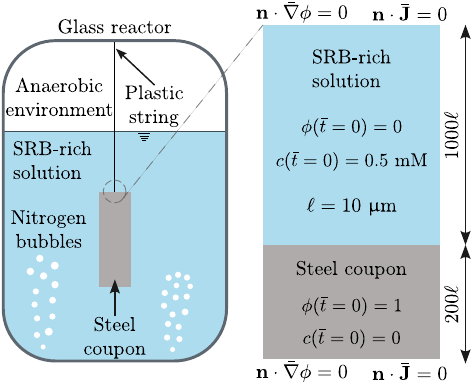}
     \captionsetup{labelfont = bf, justification = raggedright}
    \caption{Schematic illustration of the experimental setup from Ref. \cite{Peng1994} and the corresponding one-dimensional computational domain.}
    \label{Fig2}
\end{figure}

The experimental setup given in Ref. \cite{Peng1994} and schematically illustrated in Fig. \ref{Fig2} is considered for model calibration. In this experiment, steel specimens were immersed in solutions containing 0.5 mM sulfate ions and 10 mL of SRB seed. The test coupons used were 25.4 mm in diameter and 1 mm thick. Their nominal composition included 0.01\% Si, 0.01\% P, 0.19\% Mn, 0.01\% S, and balance Fe (in wt.\%). The solution pH was adjusted to 7.0, and it remained approximately equal to 7.2 throughout the experiments. The sulfate concentration was also maintained constant in the bulk solution. All experiments were performed in a closed rectangular glass reactor. Pure nitrogen was bubbled through the reactor and the solution reservoir to keep an anaerobic environment. The composition of the solution and information on the preparation of SRB seeds are available in Ref. \cite{Peng1994}. The container holding the solution was significantly larger than the steel specimens. The experiment was conducted at a constant temperature of 20 $\degree$C. Pit depth measurements were recorded at 15, 30, 50, 70, 90, 110, 130, and 150 days of immersion for three steel specimens.

A schematic illustration of the experimental setup and the corresponding computational domain is depicted in Fig. \ref{Fig2}. Since only pit depths were measured in the experiment, the simulation is simplified and considers a one-dimensional computational domain. The metal used in the simulation is 1 mm long. The size of the liquid phase is significantly larger than that of the metal phase to mimic the experimental setup. The initial sulfate ion concentrations are set to 0 mM in the metal phase and 0.5 mM in the liquid phase to match the experimental conditions. The boundary conditions for the phase-field and diffusion equations ($\mathbf{n} \cdot \bar{\nabla} \phi = 0$ and $\mathbf{n} \cdot \bar{\mathbf{J}} = 0$)  are prescribed on both edges of the computational domain. These conditions imply that no diffusion occurs across the domain boundary. The role of mechanical fields in influencing corrosion kinetics is not accounted for in this simulation, and hence, the mechanical equilibrium equation is not considered.

\subsubsection{Choice of parameters and model predictions} \label{sec312}

The material, solution and model parameters are given in Table \ref{table1}. The density of SRB and the maximum utilisation rate of the sulfate are set to $N = 6\times10^{10}$ cells/m$^3$ and $q = 2.08\times10^{-4}$ mol/(kg$\cdot$s), reflecting the experimental conditions \cite{Peng1994}. The Monod half-velocity coefficient $K_\mathrm{m}$ and the maintenance-respiration coefficient $b$ are used to describe freshwater environments and estimated to $K_\mathrm{m} = 4\times10^{-3}$ mol/m$^3$ \cite{Peng1994, Ingvorsen1984b, AlDarbi2005} and $b=10^{-8}$ 1/s \cite{Goldhaber1982}. The true yield $Y$ is set to $Y = 4.1\times10^{-2}$ kg/mol \cite{GANTZER2003}. The bacterial density in the biofilm for the freshwater environment is estimated as $c_b = 1\times 10^{-3}$ kg/m$^3$, based on experimental cell counts reported in Ref. \cite{Peng1994}, to represent an effective biofilm density during early-stage attachment. The effective stored cell energy is estimated from the calorific content of the cell \cite{Wang1976} and set to $\Psi = 3\times10^{-10}$ J/cell. The interface thickness is taken to be significantly smaller than the sample dimensions and set to $\ell = 10$ $\upmu$m, providing a suitable compromise between geometric resolution and computational efficiency. The interfacial energy is set to $\Gamma = 2.10$ J/m$^2$ \cite{Kanhaiya2014}. The phase-field model parameters $\omega$ and $\kappa$ are computed using $\ell$ and $\Gamma$ as outlined in Section \ref{sec25p}. In addition to these parameters, the interfacial mobility $L_0$ for stress-free corrosion in Eq. (\ref{eqn10}), which controls interface motion and determines the rate-limiting process governing the system (Eqs. (\ref{eqn15}) and (\ref{eqn16})), needs to be adopted. The order of magnitude of this coefficient is obtained from the sharp interface analysis in \ref{appendixB} and then tuned to quantitatively reproduce the experimental measurements \cite{Peng1994}.

\begin{table}[t]
\centering
\begin{tabular}{ l l l } 
 \hline
 Quantity & Value & Unit\\
\hline
Absolute temperature $T$ & 293.15 & K \cite{Peng1994}\\
Molar volume of the metal $V_\mathrm{m}$ & $6.93 \times10^{-6}$  & m$^3$/mol \cite{BS2005}\\
Diffusion coefficient of sulfate ions in the electrolyte $D^\mathrm{l}$ & 7.69$\times 10 ^{-10}$ & m$^2$/s \cite{Peng1994}\\
Diffusion coefficient of sulfate ions in the solid phase $D^\mathrm{s}$ & $D^\mathrm{l}\times10^{-3}$ & m$^2$/s\\
Interfacial energy $\Gamma$ & 2.10 & J/m$^2$ \cite{Kanhaiya2014}\\
Interface thickness $\ell$ & 10 & $\upmu$m\\
Energy barrier height $\omega$ & $3\Gamma/(4 \ell)$ & J/m$^3$ \cite{KOVACEVIC2020}\\
Gradient energy coefficient $\kappa$ & $3\Gamma \ell/2$ & J/m \cite{KOVACEVIC2020}\\
Effective stored cell energy $\Psi$ & $3\times10^{-10}$ & J/cell \cite{Wang1976}\\
Interfacial mobility parameter $L_0$ & $2.10\times10^{-12}$  & m$^3$/(J$\cdot$s)\\
Density of SRB cells $N$ & $6\times10^{10}$  & cell/m$^3$ \cite{Peng1994}\\
Monod half-velocity coefficient $K_\mathrm{m}$ & $4\times10^{-3}$  & mol/m$^3$ \cite{Peng1994, Ingvorsen1984b, AlDarbi2005}\\
True yield $Y$ & $4.1\times10^{-2}$ & kg/mol \cite{GANTZER2003}\\
Maintenance-respiration coefficient $b$ & $1\times 10^{-8}$ & 1/s \cite{Lin2001, GANTZER2003}\\
Maximum utilisation rate of the sulfate $q$ & $2.08 \times 10^{-4}$ &  mol/(kg$\cdot$s) \cite{Peng1994}\\
Bacterial density $c_b$ & $1 \times 10^{-3}$ &  kg/m$^3$ \cite{Peng1994}\\
\hline
\end{tabular}
\captionsetup{labelfont = bf,justification = centering}
\caption{Parameters used to calibrate the model.}
\label{table1}
\end{table}

\begin{figure}[h!]
    \centering
    \includegraphics[width = 8 cm]{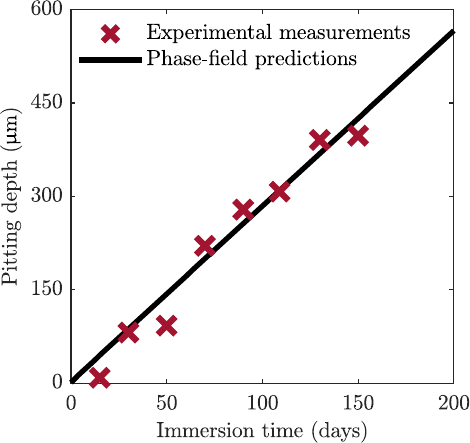}
     \captionsetup{labelfont = bf, justification = raggedright}
    \caption{Comparison between experimental measurements \cite{Peng1994} and phase-field predictions for pitting depth as a function of immersion time.}
    \label{Fig3}
\end{figure}

The model predictions in terms of pitting depth as a function of immersion time, along with the experimental data, are shown in Fig. \ref{Fig3}. A very good agreement is attained, overall. The experimental data indicate that corrosion progressed slowly in the early stage of exposure. This may be interpreted as the time required for biofilm formation and SRB acclimation to their new environment. Subsequently, corrosion progressed, leading to more severe pitting. The same trend in results is reproduced by the present framework, except for the initial immersion stage. This discrepancy between model predictions and experimental measurements at very early immersion times is expected, as biofilm formation and SRB adaptation are not considered in the current framework. That is, the present model assumes that corrosion begins immediately upon immersion without a lag period for biomass attachment and development. Nevertheless, the current anaerobic MIC model appears to be reasonably successful at predicting pitting depth influenced by SRB in sulfate-containing aqueous systems. 
This agreement between the experiments and the model prediction indicates that the value of the interfacial mobility parameter $L_0 = 2.10\times10^{-12}$ m$^3$/(J$\cdot$s) is appropriately chosen to accurately reproduce the experimental data on pitting depth. The sharp interface analysis (\ref{appendixB}) estimates a slightly lower value of $L_0 = 1.13\times10^{-12}$  m$^3$/(J$\cdot$s). The small discrepancy between these two values is expected and higher mobility is required in dynamic simulations to account for transient diffusion gradients across the interface thickness. 

For the magnitude of $L_0$ adopted and the values listed in Table \ref{table1}, one can show that $\tau\ll 1$ and $D_a \ll1$, indicating that the system operates in an interface-controlled and biokinetically limited regime. In this regime, sulfate transport is sufficiently rapid to maintain a near-uniform concentration in the electrolyte. The corrosion rate is then primarily governed by the availability of electrons, which is dictated by microbial activity. This observation explains the linear increase in pitting depth with immersion time observed in Fig. \ref{Fig3}, as the interface velocity is controlled by reaction kinetics rather than by transport limitations. This further means that in this regime, the results are weakly sensitive to variations in the diffusion coefficient $D^l$, as diffusion does not limit the overall corrosion rate under these conditions.

\subsection{Sensitivity analysis} \label{sec32}

The sensitivity of the model response to variations in biological input parameters is further investigated using the computational setup from the previous example. The parameters $K_\mathrm{m}$, $q$, $b$, $Y$, $N$, and $\Psi$ are associated with microbial kinetics and energetics and exhibit variability due to differences in SRB species, environmental conditions, and experimental measurement challenges \cite{Peng1994, AlDarbi2005, Sorensen2012, Haile2015, Wang1976, Dong2023, Saxena2025}. Changes in $K_\mathrm{m}$, $q$, $b$, $Y$ modify the threshold concentration $S_\mathrm{min}$ and shift the effective activation of microbial activity. Through this dependence, they modify both the thermodynamic driving force for interface motion $\Delta g $ (\ref{appendixB}) and the Damk\"ohler number $D_a$, altering the balance between microbial sulfate consumption and sulfate supply from the bulk electrolyte. Variations in $N$ and $\Psi$ influence only $\Delta g $ through the energy parameter $\Lambda$, without directly influencing the reaction kinetics. In each sensitivity case study, five parameters are held constant, and their values are based on Table \ref{table1}. The corrosion kinetics are evaluated by varying the sixth parameter by $\pm$ 50\% of its reference value, listed in Table \ref{table1}. All simulations are performed at a fixed bulk sulfate concentration of 0.5 mM, and the results are compared with experimental measurements.

In addition, the effect of bacterial density in biofilm $c_b$, which influences only the rate of sulfate consumption, is also examined. Although $c_b$ and $N$ can be related through the average cell mass, they are not necessarily equivalent. $N$ represents the effective number of SRB cells per unit volume, whereas $c_b$ represents biomass density. Accordingly, different values of $c_b$ may correspond to the same value of $N$ if the average cell mass varies. Therefore, $N$ and $c_b$ are treated as distinct effective parameters in the sensitivity analysis. To this end, the bacterial biofilm density is increased by factors of 50 and 100 to represent a parametric increase in attached biomass, ranging from early-stage attachment to more developed biofilms. This variation serves as a parametric representation of biomass evolution and maturation over time \cite{Costerton1995}. A longer immersion time is considered in this simulation to allow the system to reach a regime in which sulfate consumption becomes significant and the corrosion rate approaches a steady value. Sulfate diffusivity and bulk sulfate concentration are excluded from this sensitivity analysis. As discussed in the previous section, under the parameter adopted in Table \ref{table1}, the system is in an activation-controlled regime ($\tau \ll1$ and $D_a \ll 1$), where corrosion kinetics are governed by interface reactions rather than transport of SO$_{4}^{2-}$ ions. The sensitivity to the bulk sulfate concentration is omitted due to the linear dependence of the thermodynamic driving force for interface motion on sulfate concentration (\ref{appendixB}), which does not alter the qualitative behaviour of the system.

\begin{figure}[htp!]
    \centering
    \includegraphics[width = 16 cm]{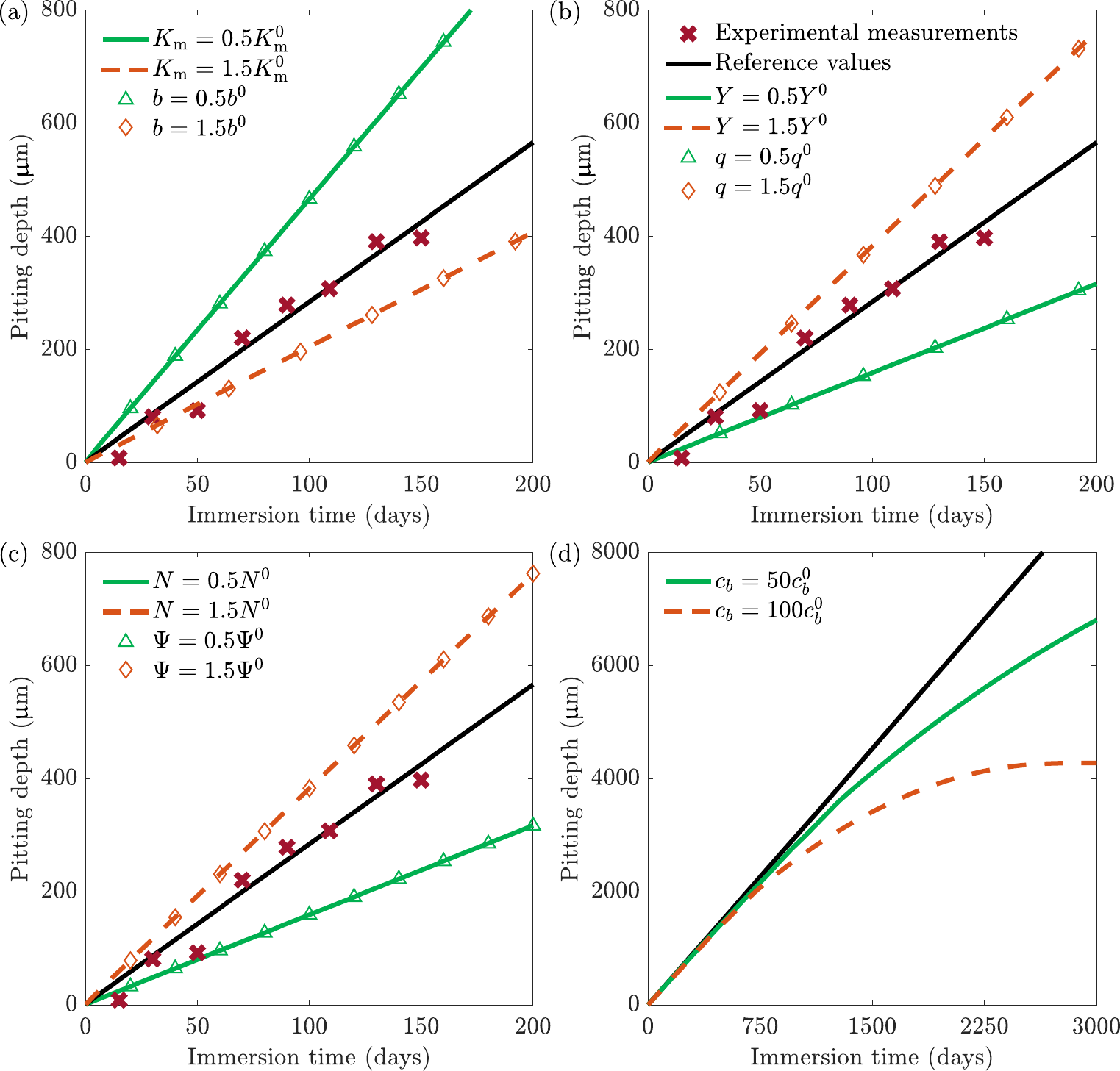}
     \captionsetup{labelfont = bf, justification = raggedright}
    \caption{Sensitivity analysis. Dependence of pitting depth on variations in the (a) Monod half-velocity $K_\mathrm{m}$ and maintenance-respiration coefficients $b$, (b) true yield of bacterial mass $Y$ and maximum specific rate of metal substrate utilisation by SRB $q$, (c) bacteria density $N$ and effective energy $\Psi$, and (d) bacterial density in the biofilm $c_b$ as a function of immersion time. Reference values for $K_\mathrm{m}^0$, $b^0$, $Y^0$, $q^0$, $N^0$, $\Psi^0$, and $c_b^0$ correspond to data in Table \ref{table1} used for model calibration in Section \ref{sec31}.}
    \label{Fig4}
\end{figure}

Figure \ref{Fig4} summarises the results for all varied parameters, along with the experimental measurements and model predictions based on the data in Table \ref{table1}. As shown in Fig. \ref{Fig4}(a), increasing the parameters $K_\mathrm{m}$ and $b$ (i.e., decreasing $\Delta g$ and $D_a$) results in lower pitting depths, which is in agreement with expectations. A 50\% reduction in $K_\mathrm{m}$ and $b$ increases pitting depth by approximately 64\%, while a 50\% increase reduces it by 28\%. The parameters $Y$ and $q$ have a net positive influence on corrosion kinetics. Increasing either parameter enhances the thermodynamic driving force $\Delta g$, leading to accelerated interface motion, Fig. \ref{Fig4}(b). At the same time, both parameters increase $D_a$, which promotes faster sulfate consumption and tends to reduce corrosion kinetics by limiting sulfate availability at the interface. However, within the range of variation considered here, the increase in $\Delta g$ dominates over the depletion effect associated with $D_a$, resulting in an overall acceleration of corrosion. Although $D_a$ exhibits quadratic dependence on $q$, the $\pm 50\%$ variation around the reference value (Table \ref{table1}) is insufficient to produce a significant difference between the responses to changes in $Y$ and $q$. Quantitatively, a 50\% decrease in either parameter reduces the pitting depth by approximately 44\%, whereas a 50\% increase leads to an enhancement of about 35\%.

Both the density of SRB cells $N$ and effective energy $\Psi$ exhibit a positive correlation with pitting depth (Fig. \ref{Fig4}(c)) as they define the driving force through the dimensionless number $\Lambda$. Increasing either parameter elevates the energy level of the system, thereby accelerating interface motion and enhancing corrosion kinetics. For instance, a 50\% increase in $N$ or $\Psi$ results in an average increase of 35\% in pitting depth, whereas a 50\% decrease leads to a reduction in corrosion kinetics of about 41\%. The trend is consistent with expectations, as both $N$ and $\Psi$ steepen the chemical free energy density and increase the chemical driving force for interface evolution. Although $N$ and $\Psi$ affect $\Lambda$ in a similar manner, small differences in pitting depth arise due to the involvement of $N$ in the reaction kinetics through the ratio $D_a$. In the present analysis, it is assumed that changes in $N$ proportionally affect the bacterial density $c_b$.  The observed increase in corrosion kinetics with higher SRB density is consistent with the literature \cite{FAN2021, Wang2026}.

The role of bacterial biofilm density $c_b$ in pitting kinetics is illustrated in Fig. \ref{Fig4}(d). Unlike the other biological parameters, $c_b$ affects only $D_a$ without altering the thermodynamic driving force $\Delta g$ or the threshold concentration $S_\mathrm{min}$. Its influence is therefore purely kinetic, acting through the rate of sulfate consumption. Since corrosion is directly coupled to sulfate reduction, the availability of sulfate governs the duration and rate of corrosion. At low and moderate values of $c_b$ (e.g., early-stage of biomass attachment), sulfate remains sufficiently available at the interface. In this case, the process proceeds at the maximum biological potential set by the utilisation rate $q$. As $c_b$ increases (e.g., mature biomass), sulfate is consumed more rapidly, and the system transitions toward a transport-limited regime, in which sulfate supply from the bulk becomes insufficient to sustain microbial activity at the interface. This transition is evident when $c_b$ is increased by a factor of 50 relative to the reference value. In this case, pitting depths initially follow the reference trend but slow down at longer immersion times due to sulfate depletion near the interface. As a result, the local sulfate concentration in the vicinity of the corroding interface becomes the rate-limiting factor and the system progressively shifts toward transport-controlled behaviour. A further increase in $c_b$ amplifies this effect. Sulfate is rapidly depleted, leading to a significant reduction in microbial activity. Consequently, corrosion kinetics are strongly suppressed, and the pitting depth approaches a plateau after approximately 2500 days of immersion. This plateau reflects the exhaustion of the electron acceptor (sulfate), beyond which SRB cells remain inactive and further corrosion cannot proceed.

\section{Results: application to engineering-relevant MIC problems}
\label{sec4}

The proposed framework is applied to assess MIC in offshore wind monopile foundations, a representative structural system where MIC is a critical design and integrity factor \cite{Liu2023}. The study investigates how the underlying microstructure of the material and cathodic protection (CP), a dominant protection measure in offshore construction, influence the initiation and progression of pitting and stress-assisted corrosion.

\subsection{Geometry and environmental conditions} \label{sec41}

The system considered is an offshore wind monopile foundation with a diameter of 10 m and a total length of 85.5 m. Approximately 45 m of the structure is embedded in the seabed, while 40 m is submerged in seawater, with an additional transition piece above the mean water level \cite{ERDOGAN2021}. The submerged section is protected using sacrificial Al–Zn–In alloy anodes with an electrochemical capacity of 2000 Ah/kg and an equilibrium potential of $E_\mathrm{eq}^\mathrm{anode} = -1.05$ V vs. Ag/AgCl, a preferred material for offshore monopiles as per the DNV-RP-B401 standard \cite{DNV2021}. The design employs slender stand-off anodes measuring 2 m in length and 0.22 m in diameter. They are arranged circumferentially around the monopile and distributed in six equally spaced rings along the submerged section to achieve full polarization \cite{ERDOGAN2021}. Fourteen anodes are used per ring. The transition piece is assumed to be coated and protected with the same sacrificial anodes as the submerged region, thereby fully preventing corrosion. The buried section is assumed to be uncoated and exposed to the aggressive bacterial environment. The structure is assumed to be constructed from S355 structural steel.

\begin{figure}[h!]
    \centering
    \includegraphics[width = 16 cm]{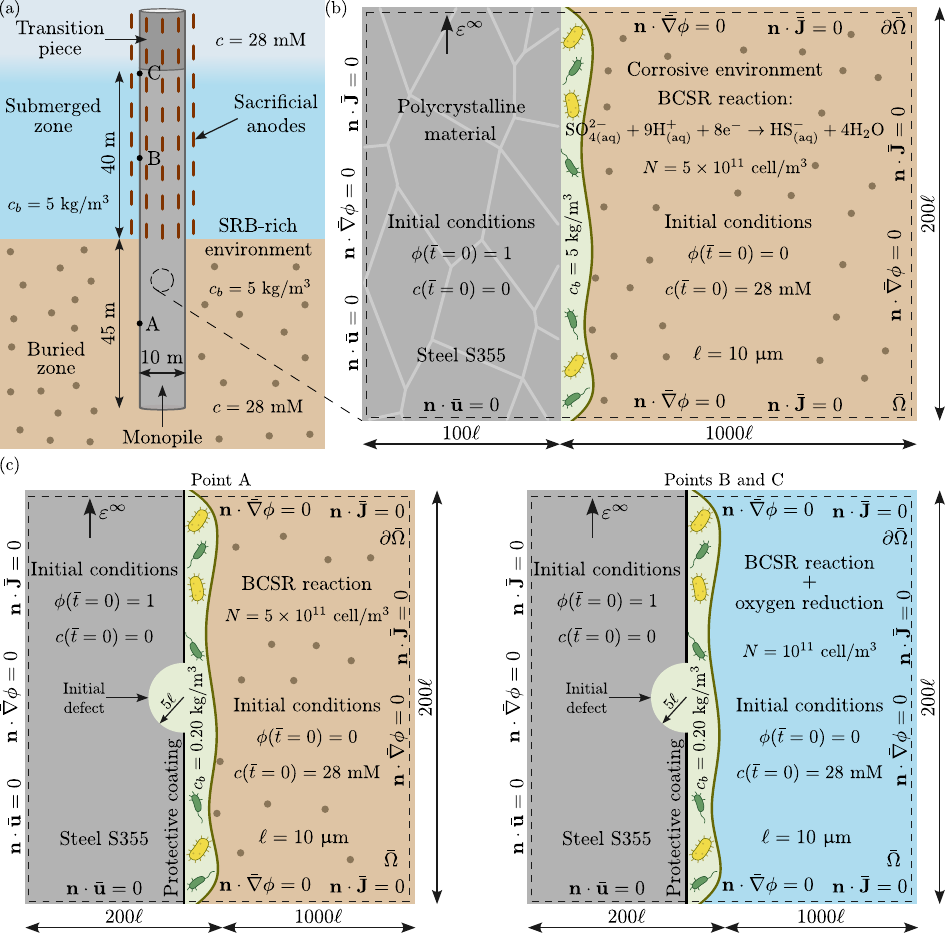}
     \captionsetup{labelfont = bf, justification = raggedright}
    \caption{Application case study. (a) Schematic illustration of an offshore wind monopile foundation with sacrificial anodes. Computational domains with initial and boundary conditions for (b) microstructure-sensitive simulations of corrosion in the buried zone and (c) structural scale simulations at three representative points.}
    \label{Fig5}
\end{figure}

The environmental conditions are selected to represent typical marine and seabed environments. Sulfate concentration $c$ and Monod half-velocity coefficient $K_\mathrm{m}$ in both submerged and buried regions are set to $c =$ 28 mM \cite{Jamieson2013} and $K_\mathrm{m} = 0.2$ mol/m$^3$ \cite{Ingvorsen1984, AlDarbi2005}. The maintenance-respiration coefficient $b$ is estimated from \cite{GANTZER2003} and is set to $b=1.20\times10^{-7}$ 1/s for both submerged and seabed zones. The density of SRB cell $N$ and the diffusion coefficient of sulfate ions in the environment $D^\mathrm{l}$ are made different in the two zones to reflect variations in bacterial availability and sediment properties. $N = 10^{11}$ cell/m$^3$ \cite{Overmann2016} and $D^\mathrm{l} = 9\times 10^{-10}$ m$^2$/s \cite{YUANHUI1974} are specified in the submerged zone. The corresponding values in the buried region are set to $N = 5 \times 10^{11}$ cell/m$^3$ \cite{Overmann2016} and $D^\mathrm{l} = 5\times 10^{-10}$ m$^2$/s \cite{YUANHUI1974}. The diffusion coefficient is reduced in the buried zone due to the tortuosity of the sediment. The bacterial biofilm density is estimated to $c_b = 5$ kg/m$^3$, which reflects the increased biomass density of a partially developed biofilm \cite{Peng1994, AlDarbi2005, Haile2015}. The other material properties, microbial parameters, and model constants follow those used in the previous example and are listed in Table \ref{table1}. A graphical illustration of the foundation, the corrosive environment, and the sacrificial anode layout is portrayed in Fig. \ref{Fig5}(a).

\subsection{Representative case studies} \label{sec42}

The analysis is divided into two case studies that address distinct length scales and operating conditions. In the first case study, corrosion evolution is examined in the buried zone without the application of CP or other mitigation measures, representing a worst-case scenario. The focus is on elucidating the role of the underlying microstructure of the material in pitting and stress-assisted corrosion under purely microbially driven conditions, Fig. \ref{Fig5}(b). In this case, microbial activity is assumed to be the only cathodic process governing corrosion.

The second case study investigates the effect of CP on corrosion evolution and defect propagation at the structural scale. In this scenario, CP is applied to the entire monopile using sacrificial anodes, protecting both submerged and buried sections. In addition to SRB-induced corrosion, the oxygen reduction reaction is included in the submerged zone to reflect realistic environmental conditions. For simplicity, the computations are conducted at three representative locations along the monopile: the midpoint of the buried section (point A), the midpoint of the submerged region (point B), and a point close to the top of the monopile (point C), as depicted in Fig. \ref{Fig5}(a). Microstructural effects are not considered in this case. Instead, the analysis focuses on the growth of a single surface defect at these representative points, Fig. \ref{Fig5}(c).

\subsection{Case study 1: Microstructure sensitive phase-field simulations of MIC}  \label{sec43}

\subsubsection{Pitting corrosion} \label{sec431}

Various microstructures with different average grain sizes are considered. Variations in average grain size are introduced to mimic different structural regions of the monopile, e.g., the base metal, heat-affected zones, and weld metal areas. Three different microstructures with average grain sizes of 30 $\upmu$m, 40 $\upmu$m, and 60 $\upmu$m are considered to cover the variations in grain size frequently reported in the literature \cite{Elahi2025, Lehto2022, Lehto2016}. The computational domain and the underlying initial and boundary conditions are depicted in Fig. \ref{Fig5}(b). The computations are conducted in two- and three-dimensional domains. The size of the corrosive environment is significantly larger than the metal geometry to more closely resemble realistic environmental conditions. The material properties and content of the surrounding environment are the same as those in Table \ref{table1} and Section \ref{sec41}.

\begin{figure}[h!]
    \centering
    \includegraphics[width = 16 cm]{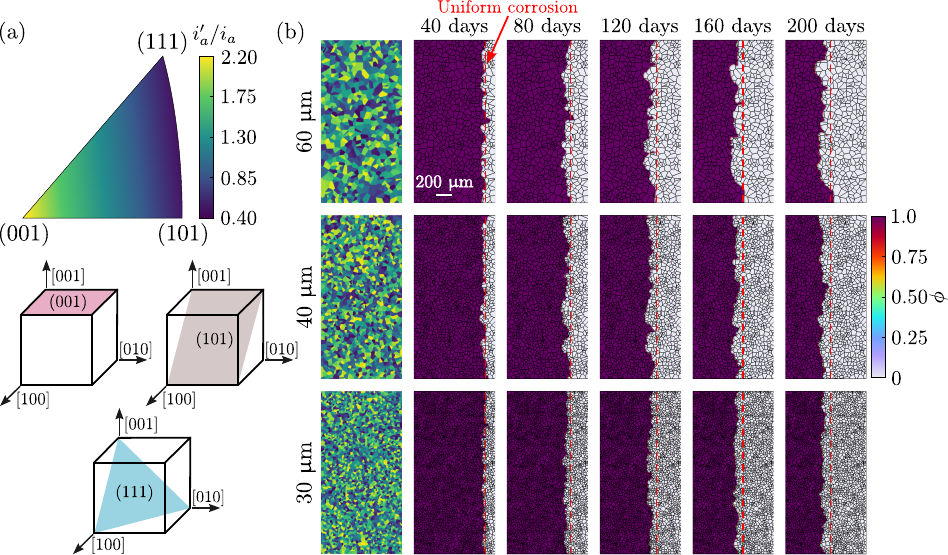}
     \captionsetup{labelfont = bf, justification = raggedright}
    \caption{Sensitivity of corrosion kinetics on microstructure. (a) Dependence of corrosion current density on crystallographic orientation. (b) Corrosion evolution in the buried zone in the absence of CP for microstructures with an average grain size of 30 $\upmu$m, 40 $\upmu$m, and 60 $\upmu$m as a function of immersion time. The red lines correspond to uniform corrosion. The initial surrounding corrosive environment is not shown in the plots. The scale bar for all plots is 200 $\upmu$m.}
    \label{Fig6}
\end{figure}

To account for crystallographic anisotropy, the phase-field mobility is made grain-dependent. Using the proportionality between phase-field mobility and anodic current density (see \ref{appendixB}), the mobility assigned to each grain is defined as: $L_0^\prime = i_a^\prime/i_aL_0$, where $i_a^\prime$ is the grain-dependent anodic current density and $i_a$ the average macroscopic value. Experimental studies show that $i_a^\prime$ varies with crystallographic orientation \cite{JIN2022}. Following reported orientation-dependent deviation factors for the (001), (111), and (101) lattice planes, $m_{(001)}=2.56$, $m_{(111)}=0.31$, and $m_{(101)}=0.13$ \cite{FUSHIMI2010}, $i_a^\prime$ is assigned to each grain from a uniform distribution with mean ($m_{(001)} + m_{(101)} + m_{(111)}$)/3 and width ($m_{(001)} + m_{(101)}$)/2. The uniform distribution is adopted in the absence of texture information to represent orientation-dependent electrochemical variability without favouring any particular crystallographic orientation. The resulting map of the ratio $i_a^\prime/i_a$ is plotted in Fig. \ref{Fig6}(a). This distribution is used as a simplified stochastic representation of orientation-dependent electrochemical heterogeneity rather than a direct reconstruction of crystallographic texture. Consequently, the simulations are intended to provide insight into the role of microstructural heterogeneity in corrosion evolution and not to reproduce the behaviour of a specific experimentally characterised microstructure.

\begin{figure}[h!]
    \centering
    \includegraphics[width = 16 cm]{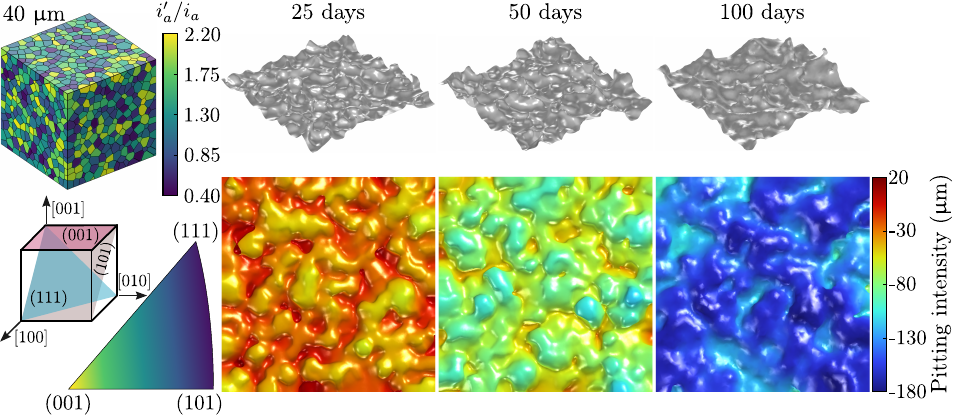}
     \captionsetup{labelfont = bf, justification = raggedright}
    \caption{Corrosion evolution in the buried zone in the absence of CP. Pitting corrosion for 3D microstructures with an average grain size of 40 $\upmu$m as a function of immersion time. Pitting intensity is defined as the difference in corrosion depth between pitting and uniform corrosion.}
    \label{Fig7}
\end{figure}

The phase-field contours for the three average grain sizes as a function of immersion time are presented in Fig. \ref{Fig6}(b). The results indicate that pit initiation and progression are strongly influenced by grain size. A higher grain size results in more severe pitting corrosion, as evidenced by deeper and more irregular pit morphologies along the corroding interface. Large grains with orientations that are more susceptible to corrosion dissolve faster, thereby accelerating the local evolution of the corrosion front. On the other hand, microstructures with smaller grain sizes promote more uniform and less localised corrosion behaviour. When microstructure-sensitive case studies are compared with the uniform corrosion (microstructure-insensitive) scenario, the corroded front advances further, resulting in greater penetration depth. The observed trend that coarse-grained microstructures exhibit greater corrosion susceptibility in the presence of SRB than fine-grained ones is consistent with experimental findings \cite{YU2023}. Although all three microstructures exhibit similar overall mass loss relative to the uniform corrosion case (plot not shown here for brevity), uneven dissolution results in severe pits, which, in the presence of mechanical loading, affect corrosion kinetics locally, as discussed further below in Section \ref{sec432}.

A three-dimensional simulation is conducted to further quantify the effect of crystallographic orientation on pitting intensity. Here, pitting intensity is defined as the difference in corrosion depth between pitting (arising from the underlying microstructure) and uniform (microstructure-insensitive) corrosion. A negative pitting intensity value implies that pitting corrosion progresses more than uniform corrosion. For simplicity, only the microstructure with an average grain size of 40 $\upmu$m is analysed. The results for corroded surfaces and pitting intensity are presented at distinct time intervals in Fig. \ref{Fig7}. The findings demonstrate that pitting intensity is less pronounced at early immersion times and increases with immersion duration, leading to greater deviations in corrosion dynamics than in the uniform corrosion case. The simulation shows that after 100 days of immersion, pitting corrosion surpasses uniform corrosion by 180 $\upmu$m. The results also reveal the complex interconnectivity between pits and their shapes, features not visible in 2D simulations.

\subsubsection{Stress-assisted corrosion}  \label{sec432}

The same corrosive environment and microstructures described above are employed to investigate the role of mechanical fields in SRB-induced corrosion kinetics. It is assumed that the structure is subjected to a remote tensile deformation of $\varepsilon^{\infty} = 0.3 \varepsilon_y$ and $\varepsilon^{\infty} = 0.5 \varepsilon_y$. The deformation, specified as a fixed displacement boundary condition, is prescribed on the top surface and held constant throughout the simulation. The corresponding computational domain, along with initial and boundary conditions for the phase-field and diffusion equations, is illustrated in Fig. \ref{Fig5}(b). Additional boundary conditions are imposed for the mechanical equilibrium equation (\ref{eqn11}). The normal component of the displacement vector $\mathbf{u}$ is constrained along the vertical and horizontal edges ($\mathbf{n}\cdot\bar{\mathbf{u}} = 0$) while the remote tensile deformation $\varepsilon^{\infty}$ is enforced on the top surface.

The material properties, the phase-field parameters $\omega$ and $\kappa$, and the stress-free microstructure-sensitive interfacial mobility $L_0^\prime$ are the same as those used in the previous example in Section \ref{sec431}. The effect of mechanical fields on corrosion kinetics is introduced through Eq. (\ref{eqn10}). The mechanical properties of steel S355 are adopted for the simulations, with Lam\'e elastic constants of $\lambda$ = 121.15 GPa and $\mu$ = 80.80 GPa \cite{EN1993}. The metal is modeled as an isotropic elasto-plastic solid. Plastic deformation is described using the J$_2$ flow theory \cite{Simo1998} with nonlinear isotropic hardening, characterized by a yield stress of $\sigma_y = 355$ MPa and a strain hardening exponent of $N = 0.1$ \cite{EN1993}. The yield strain $\varepsilon_y = \sigma_y/E$ is defined using a Young's modulus of $E = 210$ GPa \cite{EN1993}. These mechanical properties are assigned uniformly to all grains. Variations in mechanical properties arising from crystallographic orientation are neglected as their influence on corrosion kinetics is considered negligible relative to the electrochemical anisotropy \cite{Makuch2024}.

\begin{figure}[h!]
    \centering
    \includegraphics[width = 16 cm]{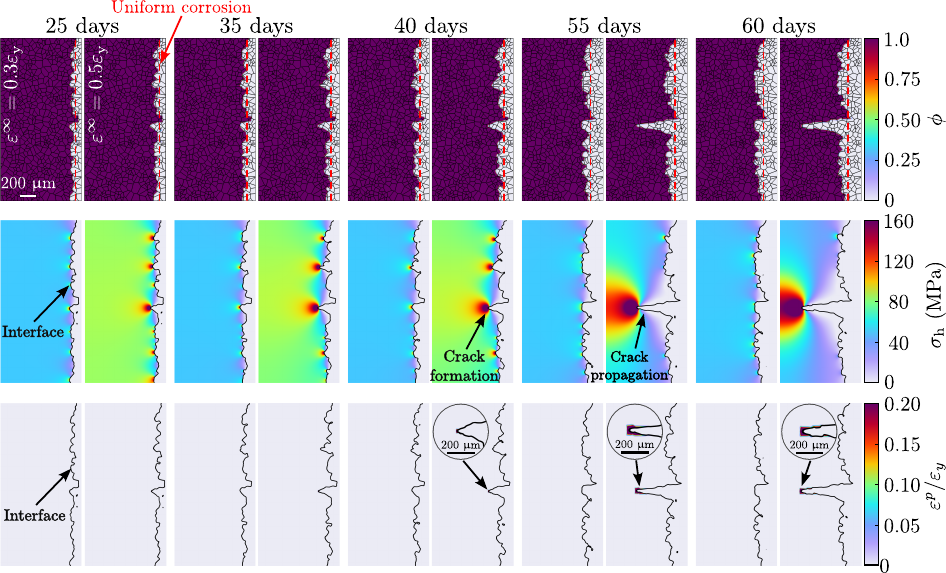}
     \captionsetup{labelfont = bf, justification = raggedright}
    \caption{Corrosion evolution in the buried zone in the absence of CP for microstructures with an average grain size of 60 $\upmu$m subjected to remote deformation $\varepsilon^{\infty}$. The red lines correspond to uniform corrosion. The initial surrounding corrosive environment is not shown in the plots. The scale bar for all plots is 200 $\upmu$m.}
    \label{Fig8}
\end{figure}

\begin{figure}[h!]
    \centering
    \includegraphics[width = 16 cm]{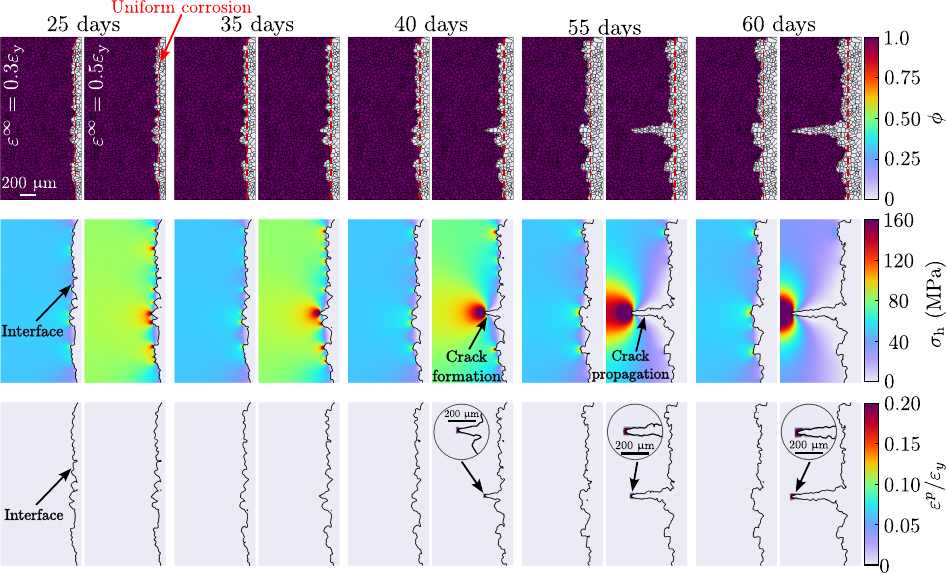}
     \captionsetup{labelfont = bf, justification = raggedright}
    \caption{Corrosion evolution in the buried zone in the absence of CP for microstructures with an average grain size of 40 $\upmu$m subjected to remote deformation $\varepsilon^{\infty}$. The red lines correspond to uniform corrosion. The initial surrounding corrosive environment is not shown in the plots. The scale bar for all plots is 200 $\upmu$m.}
    \label{Fig9}
\end{figure}

\begin{figure}[h!]
    \centering
    \includegraphics[width = 16 cm]{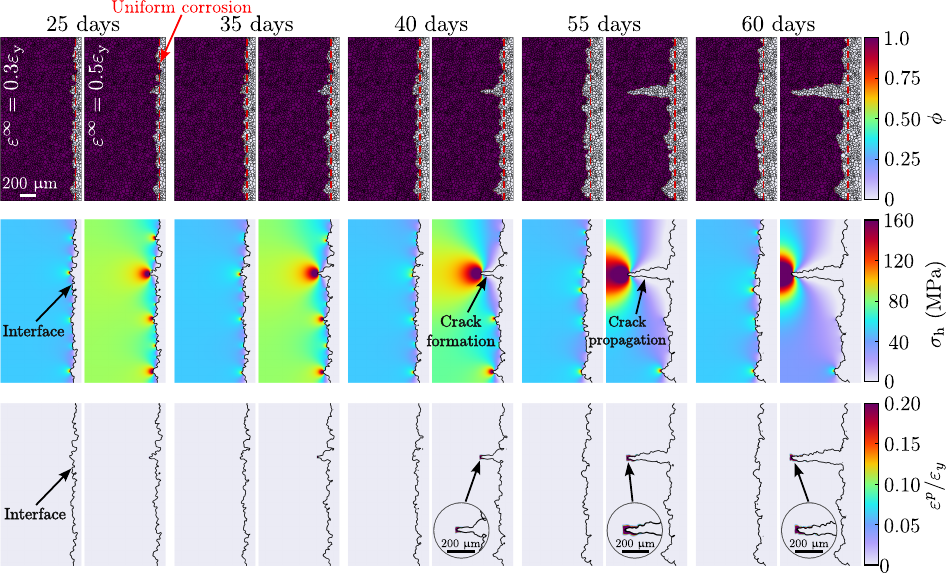}
     \captionsetup{labelfont = bf, justification = raggedright}
    \caption{Corrosion evolution in the buried zone in the absence of CP for microstructures with an average grain size of 30 $\upmu$m subjected to remote deformation $\varepsilon^{\infty}$. The red lines correspond to uniform corrosion. The initial surrounding corrosive environment is not shown in the plots. The scale bar for all plots is 200 $\upmu$m.}
    \label{Fig10}
\end{figure}

The evolution of stress-assisted corrosion in the buried zone in the absence of CP for the three microstructures is presented in Figs. \ref{Fig8}, \ref{Fig9}, and  \ref{Fig10}. Pits nucleated at the metal-environment interface act as stress concentrators in the presence of mechanical loading, accelerating dissolution locally in regions where hydrostatic stress $\sigma_\mathrm{h}$ and plastic strains $\varepsilon^{p}$ are high. The distribution of mechanical fields is highly nonuniform and depends on the evolution of the corrosion front. There are multiple small areas with high mechanical fields during early immersion times, which can potentially act as hot spots for crack initiation. The locations of these hot spots with amplified mechanical fields vary over time as the corrosion front changes shape in response to the underlying microstructure. 
As shown in Figs. \ref{Fig8}, \ref{Fig9}, and \ref{Fig10}, the low load case ($\varepsilon^{\infty} = 0.3\varepsilon_y$) does not result in crack formation, regardless of the average grain size. This load case produced zero plastic strains and low hydrostatic stresses, which are insufficient to trigger rapid local pit propagation that could serve as a site for crack initiation. On the other hand, increasing the load ($\varepsilon^{\infty} = 0.5\varepsilon_y$) activates plastic strains and higher hydrostatic stresses, leading to more rapid and severe pitting along the corroding interface. At a certain time, mechanical fields rise in a specific area, setting a precursor for a dominant location of crack initiation. After that time point, a crack forms and propagates due to high mechanical fields at the defect tip. All three microstructures exhibit crack formation after 40 days of immersion for this load case ($\varepsilon^{\infty} = 0.5 \varepsilon_y$), which then advances into rapid crack propagation. The results show that the crack path is weakly influenced by the underlying microstructure. However, its nucleation site is strongly dependent on the microstructure.

\begin{figure}[h!]
    \centering
    \includegraphics[width = 16 cm]{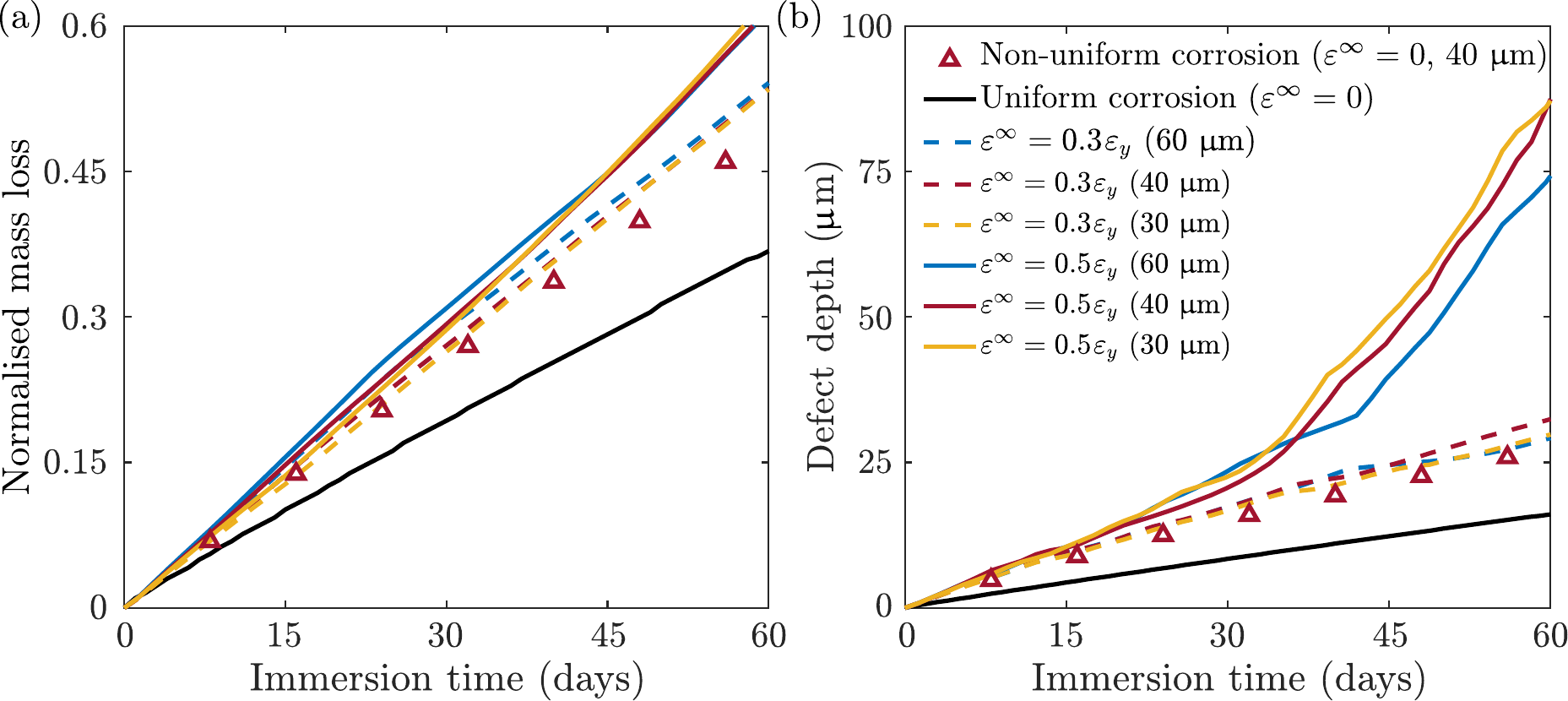}
     \captionsetup{labelfont = bf, justification = raggedright}
    \caption{Quantifying the interplay between corrosion degradation, mechanics and local heterogeneity. (a) Normalised mass loss and (b) defect depth as a function of immersion time for microstructures with an average grain size of 30 $\upmu$m, 40 $\upmu$m, and 60 $\upmu$m subjected to various remote tensile deformations $\varepsilon^{\infty}$.}
    \label{Fig11}
\end{figure}

Mechanical fields do not substantially affect the overall mass loss, as their influence is localised at the corroding interface, Fig. \ref{Fig11}(a). However, defect (crack) depth, defined here as the distance from the initial metal-environment interface to the most distant corroded point, is significantly influenced by mechanical factors. The dynamics of defect growth as a function of immersion time for the two remote tensile deformations and all three microstructures are given in Fig. \ref{Fig11}(b). The results show that more severe mechanical loading produces deeper defects that tend to transition to crack propagation at early immersion times. In the case of a low loading level ($\varepsilon^{\infty} = 0.3\varepsilon_y$), the effect of the mechanical field is not pronounced, and the three microstructures exhibit similar defect-growth kinetics, producing similar corrosion kinetics as in the case of stress-free heterogeneous corrosion. Increasing the applied load to $\varepsilon^{\infty} = 0.5\varepsilon_y$ results in much faster defect growth, even at earlier immersion times. The microstructure with an average grain size of 60 $\upmu$m exhibits the slowest defect growth, whereas the finest microstructure (average grain size of 30 $\upmu$m) exhibits the fastest defect growth under $\varepsilon^{\infty} = 0.5\varepsilon_y$. This can be explained as follows. In fine-grained microstructures, grains less susceptible to corrosion confine the corroding front to a small area before it encounters a fast-corroding grain. In the case of the coarse microstructure (average grain size of 60 $\upmu$m), grains less susceptible to corrosion impede the corrosion front over a greater distance, preventing the crack from further propagating. These findings suggest a potential trade-off between pitting resistance and crack propagation, although further studies with greater variation in microstructural morphology and grain size are indispensable to generalise this behaviour.

\subsection{Case study 2: Effect of cathodic protection on corrosion evolution} \label{sec44} 

The effect of CP on corrosion damage and defect growth is assessed by considering the entire monopile foundation, Fig. \ref{Fig5}(a). To couple structural scale electrochemical conditions with local corrosion kinetics, the analysis is divided into two steps. The first determines the spatial and temporal distribution of residual current density after the CP system is applied using an electrochemical model of the full monopile. The second addresses the assessment of corrosion damage and surface defect propagation at three representative locations along the structure using the current phase-field model. This approach constitutes a one-way multiscale coupling, in which the current density obtained in the first step is used to inform the phase-field mobility parameter $L_0$ in the subsequent corrosion simulation via the linear relationship given in Eq. (\ref{eqn9B}). The effect of the underlying microstructure is neglected in this case study.

\subsubsection{Cathodic protection calculation} \label{sec441}

The CP of the monopile is modeled by considering the entire structural geometry and surrounding environment, as shown in Fig. \ref{Fig5}(a). To conduct CP computations, the surrounding seawater and seabed are represented as cylindrical domains of sufficient size to eliminate boundary effects. The steel oxidation reaction is applied to the submerged and buried sections and is determined by following the commonly used anodic Tafel model \cite{Jones1996}
\begin{equation} \label{eqn19}
\begin{aligned}
i_a = i_0 \times 10^{\eta/A}  \quad\quad \text{with} \quad\quad \eta = \psi_\mathrm{m}-E_\mathrm{eq} - \psi,
\end{aligned}
\end{equation}
where $i_0 = 10^{-3}$ A/m$^2$ is the exchange current density, $A = 1$ V the Tafel slope, $\eta$ the overpotential, $\psi_\mathrm{m}$ the applied potential of the metal ($\psi_\mathrm{m} = 0$ V in this work), $E_\mathrm{eq} = - 0.690$ V vs. Ag/AgCl the effective macroscopic equilibrium corrosion potential for steel oxidation, and $\psi$ the solution potential in the seawater and seabed domains. The exchange current density, Tafel slope, and the equilibrium corrosion potential are estimated from Ref. \cite{KALOVELONIS2025}. The oxygen reduction reaction ($\mathrm{O_2} + 2 \mathrm{H_2O} + 4e^- \rightarrow 4 \mathrm{OH}^-$) is considered only in the submerged zone and transition piece. Under diffusion-controlled conditions typical of submerged environments, the maximum reduction current density $i_\mathrm{lim} ^\mathrm{O_2}$ is limited by the rate of oxygen transport and set to $i_\mathrm{lim} ^\mathrm{O_2} = -87\times 10^{-3}$ A/m$^2$ \cite{KALOVELONIS2025}. The current density due to the microbial activity $i_\mathrm{lim}^{\mathrm{MIC}}$ is assumed in both submerged and mud regions. It is set to a constant value of $i_\mathrm{lim}^{\mathrm{MIC}} = -20 \times 10^{-3}$ A/m$^2$, representative of the limiting current density for deep buried zones \cite{KALOVELONIS2025}. The sacrificial anode oxidation reaction is simplified with the linearised Butler-Volmer kinetics: $i_a^\mathrm{anode} = \eta/R_\mathrm{p}$ \cite{Jones1996}. Here, $R_\mathrm{p} = 0.10$ $\Omega\cdot$m$^2$ stands for the polarization resistance \cite{Calderon2022, FAROOQ2019}.

The solution potential in Eq. (\ref{eqn19}) is solved using Ohm's law 
\begin{equation} \label{eqn20}
\begin{aligned}
& - \nabla \cdot \big( \sigma_l \nabla \psi \big) = 0  \quad \text{in \,} \Omega \\
-\sigma_l \mathbf{n} \cdot \nabla \psi = i_a + i_\mathrm{lim} ^\mathrm{O_2} + i_\mathrm{lim}^{\mathrm{MIC}} \quad &\text{on \,} \partial\Omega_{s} \quad\quad \text{and} \quad\quad -\sigma_l \mathbf{n} \cdot \nabla \psi = i_a + i_\mathrm{lim}^{\mathrm{MIC}} \quad\text{on \,} \partial\Omega_{b},
\end{aligned}
\end{equation}
complemented with boundary conditions on the metal/seawater $\partial\Omega_{s}$ and metal/seabed $\partial\Omega_{b}$ surfaces. Here, $\sigma_l$ stands for the electrolyte conductivity. Far away boundaries are treated as isolated ($\sigma_l \mathbf{n} \cdot \nabla \psi = 0$). The electrolyte conductivity is set to $\sigma_l = 4.79$ S/m in seawater \cite{CRCHandbook2018} and taken in seabed as $\sigma_l = 1.40$ S/m \cite{ISO246562022}. The simulation starts with an initial value of the solution potential equal to the sacrificial anode equilibrium potential $\psi (\mathbf{x},t=0) = -E_\mathrm{eq}^\mathrm{anode}$. This computation with CP is performed over a 25-year period, which is the expected lifetime of the monopile foundation \cite{ERDOGAN2021}.  

\begin{figure}[h!]
    \centering
    \includegraphics[width = 16 cm]{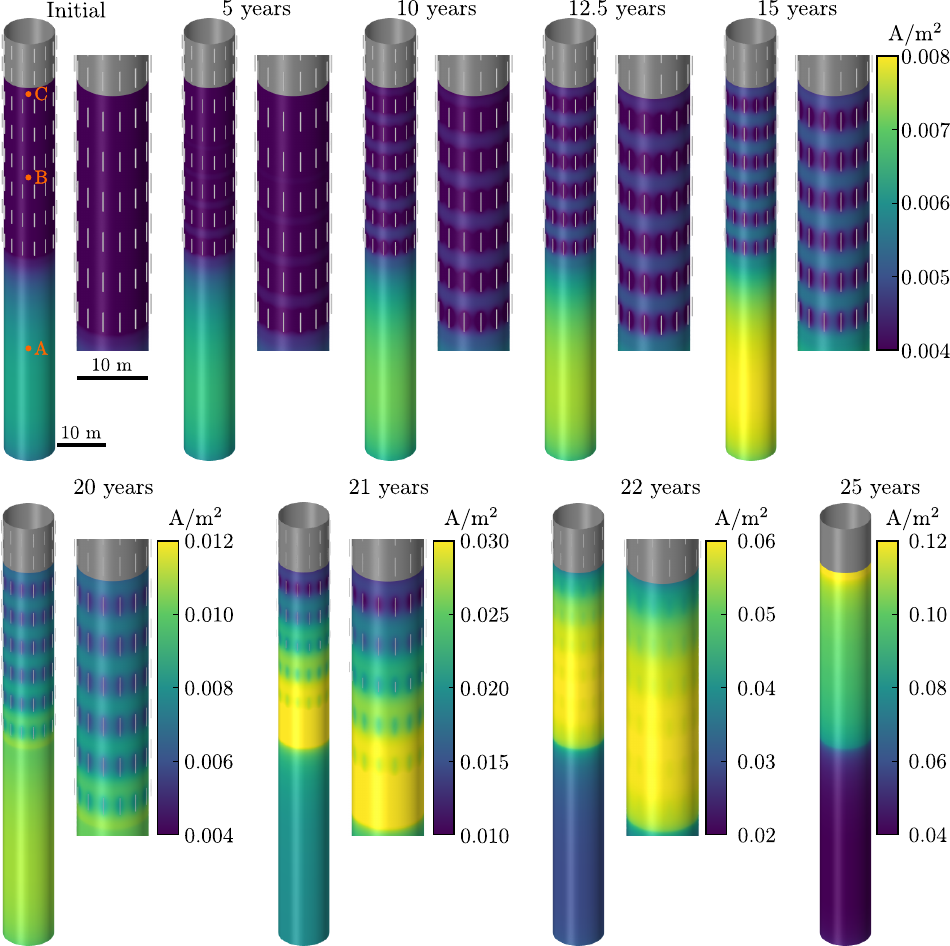}
     \captionsetup{labelfont = bf, justification = raggedright}
    \caption{Evolution of current density (A/m$^2$) for the offshore monopile foundation as a function of service time in the presence of cathodic protection. The grey bars indicate sacrificial anodes.}
    \label{Fig12}
\end{figure}

The evolution of corrosion current density along the monopile under CP conditions is depicted in Fig. \ref{Fig12}. At early stages of service, the sacrificial anodes provide effective protection, significantly suppressing the corrosion current in the submerged zone, while only a small residual current persists in the buried section. As service time increases, the sacrificial anodes progressively dissolve, reducing their size and protection efficiency. During the first fifteen years of operation (top row in Fig. \ref{Fig12}), the anodes remain largely effective, although a gradual increase in current density is observed, particularly in the buried zone and in regions between adjacent anode rings. The results indicate that anodes located near the lower part of the monopile are consumed at a higher rate, leading to earlier loss of protection in this region. Beyond fifteen years of service, the lowest ring of anodes (closest to the seabed) is fully consumed, resulting in a noticeable increase in current density that propagates from the buried region toward the submerged section. This trend becomes more pronounced after twenty years of operation (bottom row in Fig. \ref{Fig12}), as additional anode rings are depleted. The continued dissolution of sacrificial anodes leads to progressive loss of protection across both seabed and submerged zones. Complete consumption of the anodes in the submerged region occurs after approximately twenty-two years of service. Subsequent degradation of the remaining anodes in the transition region leaves the monopile fully unprotected. At this stage, the corrosion current density increases by approximately two orders of magnitude.

\begin{figure}[h!]
    \centering
    \includegraphics[width = 8.5 cm]{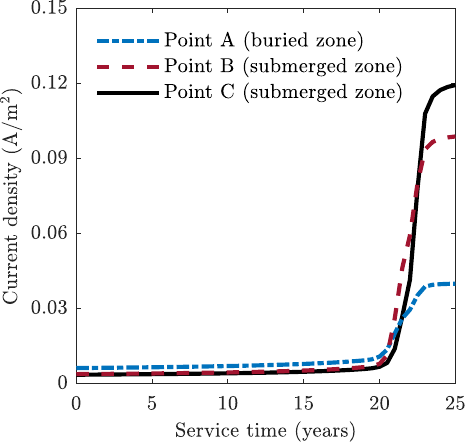}
     \captionsetup{labelfont = bf, justification = raggedright}
    \caption{Evolution of current density at the three representative points along the monopile foundation as a function of service time in the presence of cathodic protection.}
    \label{Fig13}
\end{figure}

Corrosion evolution is evaluated at the three locations along the monopile, see Fig. \ref{Fig5}(a), utilising the phase-field simulations described in the following section. The temporal evolution of the current density at these points is presented in Fig. \ref{Fig13}. At early service times, the low current densities at all locations indicate effective CP. As the sacrificial anodes dissolve, the current density increases at different rates depending on location, with the buried region exhibiting the earliest and most pronounced rise due to accelerated anode consumption in this zone. To assess the influence of CP on defect growth under pitting and stress-assisted conditions in the following section, the time-dependent current density profiles in Fig. \ref{Fig13} are fed into the present phase-field model via Eq. (\ref{eqn9B}), providing a direct one-way coupling between the local electrochemical conditions and corrosion kinetics.

\subsubsection{Phase-field simulations of defect growth} \label{sec442}

The corrosion current density obtained from the CP model in the previous section is used to inform the phase-field mobility parameter $L_0$ via Eq. (\ref{eqn9B}). Corrosion evolution at each location is modeled using two-dimensional phase-field simulations, considering the growth of a single surface defect, which is represented by a semi-circular surface pit with a radius of $5\ell$. The remaining metal-environment surface is treated as fully protected. The corresponding computational domain, along with the initial and boundary conditions, is shown in Fig. \ref{Fig5}(c). At each location, defect growth dynamics are evaluated for different mechanical loading conditions ($\varepsilon^{\infty} = 0$ and $\varepsilon^{\infty} = 0.4\varepsilon_y$) and with and without CP. In the case without CP, the current densities of $i_\mathrm{lim}^{\mathrm{MIC}}$ (applied to point A) and $i_\mathrm{lim} ^\mathrm{O_2} + i_\mathrm{lim}^{\mathrm{MIC}}$ (applied to points B and C) are used in computations.  The mechanical properties, material parameters, and model constants follow those in Table \ref{table1} and Section \ref{sec41}. The bacterial parameters are also kept the same as above, except for the bacterial density in the biofilm, which is set to $c_b=0.2$ kg/m$^3$ as an effective, time-averaged biomass density over the service life. This value lies between the low-density regime associated with early-stage attachment (Section \ref{sec3}) and higher densities representative of more developed biofilms. It is hence used to represent the cumulative effect of biofilm growth and microbial activity over extended exposure periods.

\begin{figure}[h!]
    \centering
    \includegraphics[width = 16 cm]{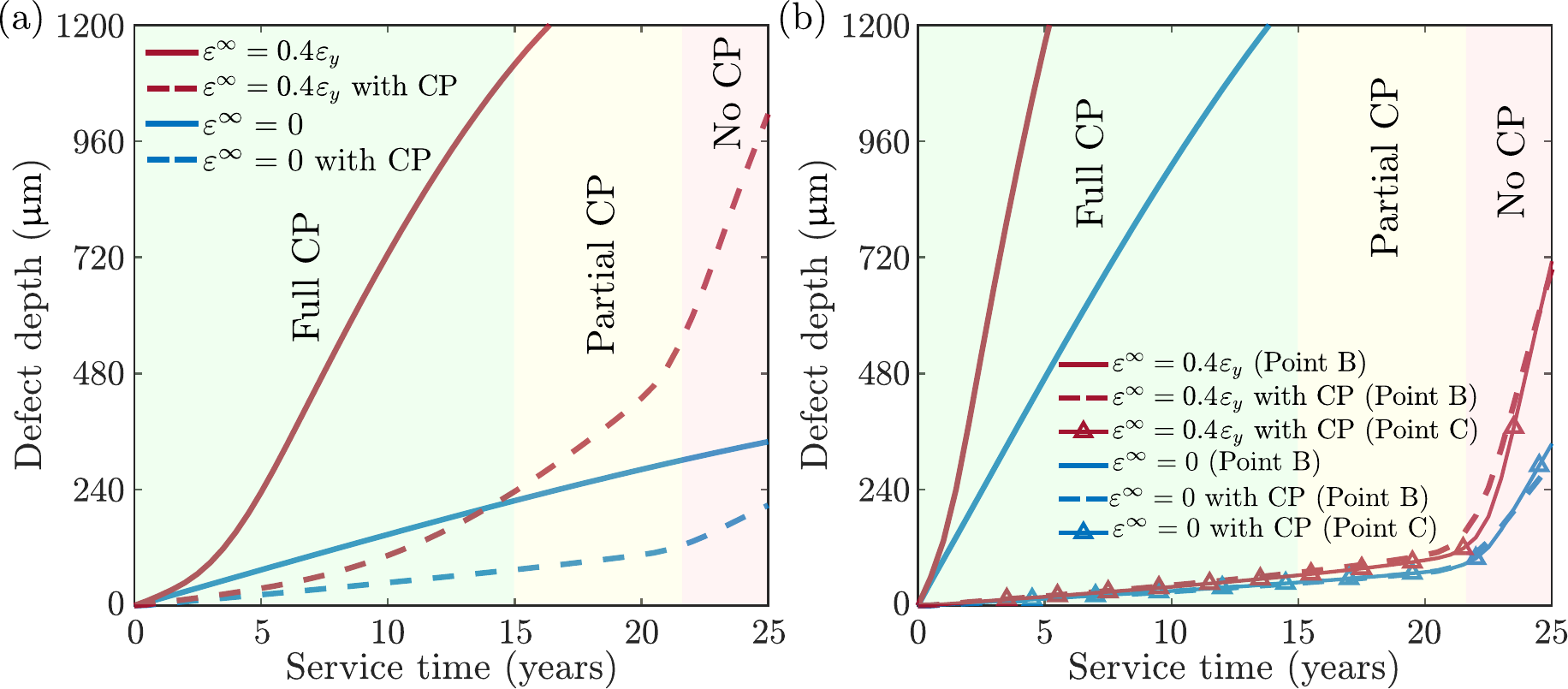}
     \captionsetup{labelfont = bf, justification = raggedright}
    \caption{Defect depth as a function of service time at (a) point A and (b) points B and C. The shaded regions indicate the different stages of cathodic protection performance: active protection (full CP), degradation (partial CP), and loss of protection (no CP).}
    \label{Fig14}
\end{figure}

\begin{figure}[h!]
    \centering
    \includegraphics[width = 16 cm]{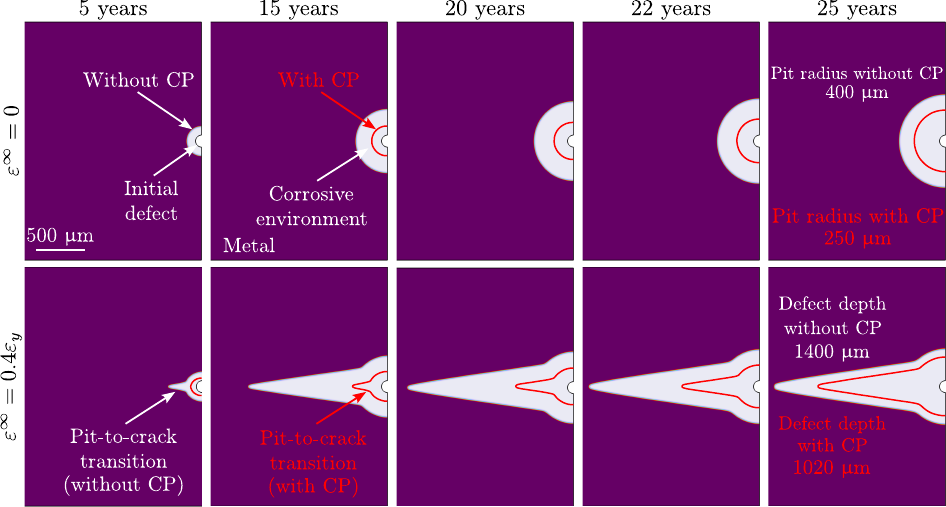}
     \captionsetup{labelfont = bf, justification = raggedright}
    \caption{Evolution of pitting ($\varepsilon^{\infty} = 0$) and stress-assisted ($\varepsilon^{\infty} = 0.4\varepsilon_y$) corrosion as a function of service time at point A. The initial surrounding corrosive environment is not shown in the plots. The scale bar for all plots is 500 $\upmu$m.}
    \label{Fig15}
\end{figure}

The temporal evolution of corrosion damage, quantified as defect depth, in the presence and absence of CP at the three representative points is depicted in Fig. \ref{Fig14}. The results demonstrate that applying CP substantially delays pitting and stress-assisted corrosion at all three points. During the initial service period (approximately fifteen years), the sacrificial anodes provide effective protection, resulting in low defect growth rates under both pitting and stress-assisted conditions. As the sacrificial anodes progressively dissolve, the corrosion kinetics increase, with the onset of accelerated degradation occurring earliest at point A in the buried zone. This behaviour is consistent with the faster anode consumption in this region. Points B and C remain effectively protected for a longer duration, with a sharp increase in defect growth observed only after complete anode depletion, which occurs approximately after twenty-two years of service, Fig. \ref{Fig13}. The visual representation of pitting ($\varepsilon^{\infty} = 0$) and stress-assisted ($\varepsilon^{\infty} = 0.4\varepsilon_y$) corrosion as a function of service time is, for brevity, only given at point A, Fig. \ref{Fig15}. Similar plots can be obtained for the other two points. In the absence of CP, stress-assisted conditions lead to rapid defect growth and early transition from pit growth to crack propagation, occurring within the first 5 years of service. This trend is reflected by the steep increase in defect depth in Fig. \ref{Fig14}. In contrast, as indicated in Fig. \ref{Fig15}, the presence of CP delays this pit-to-crack transition and reduces the rate of defect propagation, particularly during the protected service period.

From a mechanistic perspective, the effect of CP can be interpreted within the present framework as a reduction in the effective interface kinetics through its influence on the corrosion current density. Since the phase-field mobility parameter $L_0$ is proportional to the current density (\ref{appendixB}), the application of CP increases the characteristic interface time $\tau_\phi$, thereby reducing the ratio $\tau$. This shifts the system toward a more reaction-limited regime, effectively slowing interface motion without altering sulfate transport or microbial kinetics. In regions where CP is strong (e.g., location C), the reduction in $L_0$ suppresses pit growth and prevents the transition to crack propagation. Conversely, as sacrificial anodes dissolve over time, the local current density increases (Fig. \ref{Fig12} and Fig. \ref{Fig13}), reducing $\tau_\phi$ and increasing $\tau$. This leads to a progressive transition to faster corrosion kinetics, particularly in regions where sacrificial anode dissolution occurs first (location A). The same trend occurs later at points B and C due to prolonged CP effectiveness.

\section{Discussion and future work}  \label{sec5}

The present phase-field-based reaction–diffusion model provides a unified description of MIC by integrating chemical effects, microbial activity, and mechanical fields within a single computational framework. Distinguishing features of this model, compared to existing approaches in the literature \cite{Peng1994, Javaherdashti2004, AlDarbi2005, Maxwell2006, Melchers2006, Gu2009, Larsen2013, Gu2014, Haile2015, XU2016, DAWUDA2021, ANGUITA2022}, include its ability to account for (i) the time-dependent evolution of the corroding interface in arbitrary domains and (ii) the influence of mechanical fields on corrosion kinetics across distinct length scales. The former is achieved by introducing a phase-field variable that tracks the metal-environment interface, whose evolution is governed by the thermodynamic driving force derived from the variational derivative of the free energy functional, Eq. (\ref{eqn5}). The latter is incorporated following Gutman’s theory \cite{Gutman1988}, whereby mechanical fields modify the kinetics of interface motion by enhancing the phase-field mobility, Eq. (\ref{eqn10}).

The dependence of corrosion kinetics on sulfate availability is naturally embedded through the chemical free energy density. Sulfate is incorporated into the free energy functional because its availability governs the intensity of the biocatalytic cathodic sulfate reduction reaction (Eq. (\mbox{\ref{eqn1b}})) and, consequently, the rate of iron dissolution. Thus, sulfate acts both as a substrate required for SRB metabolism and as a quantity that links microbial activity to corrosion evolution. Microbial activity contributes to the thermodynamic driving force via its incorporation into the free energy functional and through the reaction term, which governs sulfate consumption and modifies the local concentration field. This dual contribution ensures that microbial processes influence both the thermodynamic driving force for interface motion and sulfate transport. The parameters governing microbial sulfate consumption are expressed in terms of experimentally measurable quantities. Environmental effects enter the formulation through the microbial and transport parameters, enabling different environments, including marine, freshwater, and landfill conditions, to be represented without changing the governing equations.

The benefits of this model are reflected in its simplicity and, at the same time, in its ability to capture key mechanisms governing MIC. Dimensional analysis in Section \ref{sec25} indicates that there are three independent parameters that govern the process: $\tau$, $D_a$, and $\Lambda$. The first one, controlled by the diffusion coefficient and phase-field mobility, distinguishes activation-controlled from diffusion-controlled behaviour. The Damk\"ohler number $D_a$ depends on biological parameters governing sulfate consumption, such as bacterial density and sulfate utilisation rate, and separates transport-limited from biokinetically limited regimes. The energy parameter $\Lambda$ compares the microbial energetics with the interfacial energy barrier, determining the strength of the bacterial contribution to interface evolution.

\subsection{Long-term MIC predictions} 

The capability of the model to perform long-term predictions is demonstrated through case studies presented in Section \ref{sec4}, considering both microstructural and structural length scales. At the microstructural scale, the model captures the influence of grain size on pitting and stress-assisted corrosion over extended time periods. The results indicate that grain size plays an important role in microbially influenced pitting and stress-assisted corrosion. While decreasing the grain size can help reduce pitting severity, it may promote faster defect propagation under loading conditions. The evolution of pitting and stress-assisted corrosion at the structural scale is analysed at representative locations along the structure. Spatially varying electrochemical conditions induced by CP are one-way coupled with the present framework, enabling corrosion predictions in realistic operating environments. Potential feedback of CP on microbial activity and interfacial chemistry is not considered. The proposed CP-MIC coupling is therefore intended to assess the influence of cathodic protection on the evolution of corrosion under prescribed microbial conditions.

The framework enables the simulation of corrosion processes over time scales that are difficult to access experimentally. In particular, it provides a computationally efficient tool for assessing the evolution of defect depth, pit-to-crack transition, and crack propagation under service conditions. This capability is especially relevant for infrastructure exposed to aggressive microbial environments, where degradation occurs over decades. However, the present formulation assumes constant microbial activity and does not explicitly account for biofilm growth and detachment dynamics, nutrient depletion, or corrosion product accumulation, all of which may influence corrosion evolution over long exposure times. Accordingly, long-term predictions are based on effective (time-averaged) microbial parameters that are representative of the prescribed environmental conditions. From an engineering perspective, the model can support service life assessment and evaluate the effectiveness of mitigation strategies (such as CP) over time. It may also serve as a complementary tool for planning experimental campaigns and for assessing the residual strength of components subjected to MIC in various environments.

\subsection{Model extension}

Although the present framework overcomes several limitations of existing MIC models \cite{Peng1994, Javaherdashti2004, AlDarbi2005, Maxwell2006, Melchers2006, Gu2009, Larsen2013, Gu2014, Haile2015, XU2016, DAWUDA2021, ANGUITA2022}, it is derived under simplified assumptions that warrant further investigation. The mechanical response of the material is described using linearised kinematics and von Mises plasticity. While this choice is primarily made for simplicity and is widely adopted in the literature \cite{LIN2021, CUI2021, Cui2023, KOVACEVIC2023}, it does not capture anisotropic deformation mechanisms. However, the present framework can readily be extended to incorporate more advanced constitutive models \cite{Song2023, Makuch2024, CHEN2026} to provide a more detailed representation of mechanical fields at the microstructural level, albeit at increased computational cost. In terms of microbial dynamics, the current formulation assumes that bacterial growth and biomass, i.e., the density of SRB cells $N$, remain constant over time, implying that cells neither reproduce nor undergo cell death. Incorporating bacterial population dynamics, for example, through additional evolution equations coupled with chemotaxis models \cite{KELLER1971}, would provide a more realistic description of biofilm development. However, such extensions introduce additional parameters that are difficult to calibrate due to limited experimental data.

The present formulation focuses on the net electrochemical effect of SRB activity. Since the biocatalytic cathodic sulfate reduction mechanism (Eq. (\mbox{\ref{eqn1b}})) provides a stoichiometric link between sulfate reduction and iron dissolution, sulfate concentration is explicitly tracked. Dissolved Fe$^{2+}$ is thus not treated as an independent field. Microbial activity is coupled to corrosion evolution through sulfate consumption, which modifies the chemical free energy density and thereby the driving force for interface motion. This simplification is adopted in the present investigation for simplicity and computational efficiency. The framework is therefore most appropriate for situations in which the overall corrosion response is governed primarily by sulfate-reduction kinetics and the net electrochemical effect of microbial activity. Explicitly resolving Fe$^{2+}$, sulfide species, pH, and electric potential fields would provide a more detailed description of aqueous chemistry, electrochemical interactions, and ionic transport. However, this would substantially increase model complexity and computational cost.

The model treats the microbial community as a single effective population, meaning that different SRB strains contribute equally to the chemical energy (Eq. (\ref{eqn6})) and to sulfate consumption through the reaction term in Eq. (\ref{eqn14}). This approach is adopted in the present work for simplicity. In reality, multiple SRB strains with different metabolic characteristics may coexist and interact. An extension to account for multiple bacterial species can be formulated by introducing species-dependent contributions to the chemical free energy and reaction kinetics
\begin{equation} \label{eqn14_a}
\begin{aligned}
& \mathcal{F}^\mathrm{chem} (\bar{c}, \phi) =  \frac{1}{2} \sum_j^m N_j \Psi_j \Big[ \bar{c} - \frac{K_\mathrm{m}^j b_j}{(Y_j q_j - b_j)c_\mathrm{ref}} \Big(1- h(\phi) \Big) \Big]^2 + \omega g(\phi),\\
&\frac{\partial \bar{c}}{\partial t} = -\sum_j^m \Big[ \nabla \cdot \mathbf{J}_j + \frac{q_j c_b^j}{c_\mathrm{ref}} \frac{\bar{c}}{\bar{K}_\mathrm{m}^j  + \bar{c}}p(\phi) \Big],\\
\end{aligned}
\end{equation}
where $N_j$, $\Psi_j$, $b_j$, $Y_j$, $q_j$, $c_b^j$, $\mathbf{J}_j$, and $K_\mathrm{m}^j$ are corresponding quantities for the $j^{\mathrm{th}}$ SRB cell type and $m$ denotes the total number of different SRB strains. The remaining expressions in Section \ref{sec24} remain the same. While this generalisation is straightforward from a modelling perspective, its practical implementation requires detailed experimental characterisation of SRB strain-specific parameters.

Finally, the present formulation does not explicitly model biofilm formation and its impact on mass transport. Instead, the biofilm is represented implicitly through a thin reaction zone in which sulfate reduction takes place. This simplification enables the focus to remain on the coupling between microbial activity, sulfate transport, electrochemistry, and mechanics, while avoiding the introduction of additional variables and parameters associated with biofilm attachment, growth, detachment, and spatial heterogeneity. The framework is thus not intended to explicitly predict biofilm morphology or its temporal evolution. As a consequence, the framework may not fully capture changes in local chemistry and mass transport conditions associated with biofilm evolution, which could influence corrosion kinetics and the extent of corrosion localisation. Incorporating biofilm growth \cite{Wanner2006} and its influence on diffusivity and local chemistry represents a natural extension of the framework and would improve the accuracy in systems where biofilm development plays a dominant role.

\section{Conclusions}  \label{sec6}

A phase-field-based reaction-diffusion corrosion model is developed to predict SRB-induced microbially influenced corrosion (MIC) in metal alloys under anaerobic conditions. The formulation is based on the biocatalytic cathodic sulfate reduction theory and integrates microbial kinetics, mass transport, and mechanical effects within a unified framework. A phase-field variable is used to track the evolution of the corrosion front, a concentration variable describes sulfate transport, and a mechano-chemical coupling is introduced by enhancing the definition of the phase-field mobility parameter to account for the interplay between corrosion kinetics and mechanical fields. In addition to uniform and pitting corrosion, the proposed framework can assess MIC under stress conditions in arbitrary geometries with no special treatment for the evolution of the corrosion front. The model assumes that environmental conditions are conducive to sustainable microbial activity, described by a Monod-type relationship for sulfate consumption. Sensitivity to microbial kinetics, energetics, and density is incorporated through both the thermodynamic driving force for interface motion and the reaction term governing sulfate consumption.

The predictive capabilities of the model are demonstrated by benchmarking against laboratory experiments and assessing the degradation of a full-scale component - namely, an offshore wind turbine monopile. The following conclusions can be drawn from the case studies:
\begin{itemize}[leftmargin=*,labelsep=0.8em]

\item The model reproduces the expected trends associated with SRB-induced MIC, including the acceleration of corrosion kinetics with increasing microbial activity and the transition from biokinetically-controlled to transport-limited behaviour at high biofilm densities.

\item The underlying microstructure of the material plays an important role in microbially influenced pitting and stress-assisted corrosion. Reducing grain size decreases pitting severity but promotes faster defect propagation under mechanical loading. While mechanical fields have a negligible impact on overall mass loss, they significantly affect local corrosion rates, which are not reflected in global corrosion response and mass loss predictions.

\item Cathodic protection (CP) using sacrificial anodes significantly alters corrosion kinetics at the structural scale. The results show that CP effectively delays pitting and suppresses crack initiation and propagation, thereby extending the service life of the structure. The degradation of CP over time leads to accelerated corrosion, highlighting the importance of accounting for time-dependent protection efficiency. These predictions reflect the influence of CP on corrosion current density and do not account for feedback of CP on microbial activity and interfacial chemistry.
\end{itemize}

The proposed framework represents a step toward predictive, mechanism-based assessment of material degradation under MIC conditions over extended time and spatial scales. It offers a computationally efficient alternative to long-term experiments and can support the design, maintenance, and life assessment of metallic systems operating in harsh microbial environments.

\section*{Acknowledgments} \label{sec7}

\noindent The authors acknowledge financial support from the EPSRC Supergen ORE Hub (Grant FF2023-1028) and UKRI’s Future Leaders Fellowship program [Grant MR/V024124/1]. The authors acknowledge the use of the University of Oxford Advanced Research Computing (ARC) facility in carrying out this work (\url{http://dx.doi.org/10.5281/zenodo.22558}).
 
\section*{Data availability} \label{sec8}

The code developed, together with example case studies and documentation, will be available at \url{https://mechmat.web.ox.ac.uk/codes} after article acceptance.

\appendix
\section{Chemical free energy density} \label{appendixA}

The chemical free energy density of the system considered (Fig. \ref{Fig1}) is decomposed into the chemical energy density associated with the concentration of sulfate ions and double-well potential energy associated with the phase-field variable. The former one is represented as the weighted sum of liquid $\mathcal{F}^\mathrm{chem}_{l}(\bar{c}_l)$ and solid $\mathcal{F}^\mathrm{chem}_{s}(\bar{c}_s)$ chemical free energy densities within the pure phases \cite{Makuch2024ECA}
\begin{equation} \label{eqn5A}
                   \mathcal{F}^{\mathrm{chem}}(\bar{c},\phi) = (1 - h(\phi))\mathcal{F}^\mathrm{chem}_{l}(\bar{c}_l) + h(\phi)\mathcal{F}^\mathrm{chem}_{s}(\bar{c}_s) + \omega g(\phi),
\end{equation}
where $\bar{c}_{l}=c_l/c_\mathrm{ref}$ and $\bar{c}_{s} = c_s/c_\mathrm{ref}$ stand for the normalised sulfate ion phase concentrations in the electrolyte and metal phases. $c_\mathrm{ref}$, $g(\phi)$, $h(\phi)$, and $\omega$ have the same definition as in Section \ref{sec2}. The chemical free energy densities within each phase are represented by parabolic functions around equilibrium concentrations as 
\begin{equation} \label{eqn8A}
                   \mathcal{F}^\mathrm{chem}_{l}(\bar{c}_l) = \frac{1}{2} N \Psi (\bar{c}_{l} - \bar{S}_\mathrm{min})^2  \quad\mathrm{}\quad \mathcal{F}^\mathrm{chem}_{s}(\bar{c}_s) = \frac{1}{2} N \Psi \bar{c}_{s}^2,
\end{equation}
where $\bar{S}_\mathrm{min} = S_\mathrm{min}/c_\mathrm{ref}$ is the normalised equilibrium sulfate ion concentration in the liquid phase. $S_\mathrm{min}$ designates the minimum (limiting) concentration of SO$_4^{2-}$ ions required for sustainable microbial activity \cite{Rittmann1980}. This value is not constant and depends on SRB strain types and environmental conditions (Section \ref{sec23}). The normalised equilibrium sulfate ion concentration in the solid phase is set to zero. $N$ and $\Psi$ are the SRB cell density and effective energy available per cell. Representing the chemical free energy density $\mathcal{F}^\mathrm{chem}_{l}(\bar{c}_l)$ with a parabolic function around the equilibrium concentration $\bar{S}_\mathrm{min}$ ensures that the sulfate concentration does not drop below the threshold value for positive microbial activity. The interfacial region is defined as a mixture of both phases with different concentrations but with the same diffusion chemical potential \cite{KKS1999}
\begin{equation} \label{eqn9A}
                   \bar{c} = (1-h(\phi)) \bar{c}_{l} + h(\phi) \bar{c}_{s}  \quad\mathrm{}\quad \frac{\partial \mathcal{F}^\mathrm{chem}_{l}(\bar{c}_{l})}{\partial \bar{c}_{l}} = \frac{\partial \mathcal{F}^\mathrm{chem}_{s}(\bar{c}_{s})}{\partial \bar{c}_{s}}.
\end{equation}
Combining Eqs. (\ref{eqn8A}) and (\ref{eqn9A}) renders the expression for the chemical free energy density of the system given in Eq. (\ref{eqn6}) in Section \ref{sec23}.

\section{Relation between phase-field mobility and corrosion current density} \label{appendixB}

To ensure that the phase-field model accurately reproduces the physical corrosion kinetics, the interfacial mobility parameter $L_0$ must be calibrated against experimental data. Different material systems and corrosive environments operate on different length and time scales, yielding kinetics that differ by orders of magnitude. Thus, selecting $L_0$ to match experimental kinetics is a trial-and-error approach. Since $L_0$ dictates the kinetic response of the diffuse interface, it is usually linearly connected to the corrosion current density $i_a$, a physical quantity that is directly related to the mass loss. Although the proportionality relation between $L_0$ and $i_a$ has been assumed in the literature without explicit form \cite{MAI2016, Cui2023, Makuch2024ECA, Ansari2018, ZENG2024}, its physical interpretation and dependence on material and model parameters remain obscure. However, a direct link between $L_0$ and $i_a$ can be estimated by mapping a 1D steady-state phase-field interface velocity $v_{pf}$ to the sharp-interface velocity $v_n$.

According to Faraday's second law, the normal velocity of the corroding front is given as
\begin{equation} \label{eqn1B}
                   v_n = i_a\frac{M}{z F \rho},
\end{equation}
where $M$ is the molar mass of the metal, $z=2$ the number of electrons exchanged in oxidation reaction (\ref{eqn1a}), and $\rho$ the density of the metal. The other symbols have the same definition as in the main text. In the phase-field model, the evolution of the corroding interface is given in Eq. (\ref{eqn9}) and can be written as \cite{ALLEN1979}
\begin{equation} \label{eqn2B}
\frac{\partial \phi}{\partial t} = -L_0\Big(\frac{\partial f^{\mathrm{chem}}}{\partial\phi} + \omega\frac{dg}{d \phi} - \kappa\nabla^2\phi \Big),
\end{equation}
where $f^{\mathrm{chem}}$ represents the part of the chemical free energy density (Eq. (\ref{eqn6})) associated with the concentration of sulfate ions. The enhanced definition of the phase-field mobility parameter with the mechano-chemical coupling (Eq. (\ref{eqn10})) is not considered in the previous equation. Assuming a 1D steady-state traveling interface profile and introducing a moving coordinate frame $\xi = x-v_{pf}t$, the previous equation becomes
\begin{equation} \label{eqn3B}
-v_{pf}\frac{d \phi}{d \xi} = -L_0\Big(\frac{\partial f^{\mathrm{chem}}}{\partial\phi} + \omega\frac{dg}{d \phi} - \kappa \frac{d^2 \phi}{d \xi^2} \Big).
\end{equation}
Multiplying the entire equation by $d\phi/d\xi$ and integrating across the infinite domain yields
\begin{equation} \label{eqn4B}
v_{pf}\int^{\infty}_{- \infty}\Big(\frac{d \phi}{d \xi} \Big)^2 d\xi= L_0\int^{\infty}_{- \infty} \Big(\frac{\partial f^{\mathrm{chem}}}{\partial\phi} + \omega\frac{dg}{d \phi} - \kappa \frac{d^2 \phi}{d \xi^2} \Big) \frac{d \phi}{d \xi} d \xi.
\end{equation}
The second term on the right-hand side is zero in the bulk phases ($g(0)=g(1) = 0$). Upon standard manipulation, the third term on the right-hand side is written as
\begin{equation} \label{eqn5B_1}
\int^\infty_{-\infty}\kappa \frac{d^2 \phi}{d \xi^2} \frac{d \phi}{d \xi} d \xi = \frac{\kappa}{2} \Big (\frac{d \phi}{d \xi} \Big)^2, 
\end{equation}
which vanishes in the pure phases ($d \phi/d \xi=0$ at $\pm \infty$). Eq. (\ref{eqn4B}) then reads
\begin{equation} \label{eqn5B}
v_{pf}\int^{\infty}_{- \infty}\Big(\frac{d \phi}{d \xi} \Big)^2 d\xi= L_0 \int^{1}_{0} \frac{\partial f^{\mathrm{chem}}}{\partial\phi} d \phi = L_0 \Delta g(\bar{c}_0),
\end{equation}
where $\Delta g(\bar{c}_0)$ is the thermodynamic driving force as a function of normalised bulk sulfate concentration $\bar{c}_0 = c_0/c_\mathrm{ref}$ across the interface (i.e., initial normalised bulk concentration). It represents the free energy density difference between the metal ($\phi = 1$) and electrolyte ($\phi = 0$) phases: $\Delta g(\bar{c}_0) = f^{\mathrm{chem}}(\phi=1,\bar{c}_0) - f^{\mathrm{chem}}(\phi=0,\bar{c}_0)$. Using the definition of $f^{\mathrm{chem}}$ from Eq. (\ref{eqn6}), the thermodynamic force can be expressed as    
\begin{equation} \label{eqn6B}
\Delta g(\bar{c}_0) =  \frac{1}{2} N \Psi \bar{c}_0^2 - \frac{1}{2} N \Psi (\bar{c}_0 - \bar{S}_\mathrm{min})^2 = N\Psi \bar{S}_\mathrm{min} \Big(\bar{c}_0 - \frac{\bar{S}_\mathrm{min}}{2} \Big).
\end{equation}
We assume that the shape of the moving interface is identical to the shape of an equilibrium (stationary) interface. At equilibrium ($v_{pf}=\Delta g = 0$), phase-field equation (\ref{eqn3B})  reduces to
\begin{equation} \label{eqn7B}
\kappa\frac{d^2 \phi}{d \xi^2} = \omega\frac{d g}{d \phi} \quad \Rightarrow \quad\frac{\kappa}{2} \Big(\frac{d \phi}{d \xi} \Big)^2 = \omega g(\phi)  \quad \Rightarrow \quad \frac{d \phi}{d \xi} = \sqrt{\frac{2\omega g(\phi)}{\kappa}}. 
\end{equation}
After substitution of the previous expression into Eq. (\ref{eqn5B}), the following relation can be obtained
\begin{equation} \label{eqn8B}
v_{pf}\int^{\infty}_{- \infty}\Big(\frac{d \phi}{d \xi} \Big)^2 d\xi = v_{pf}\int^1_0 \sqrt{\frac{2\omega}{\kappa}g(\phi)} d\phi = v_{pf}\frac{2}{3}\sqrt{\frac{2\omega}{\kappa}} = L_0 \Delta g(\bar{c}_0).
\end{equation}
Using the relation $\ell = \sqrt{\kappa/2\omega}$, the steady-state phase-field interface velocity becomes
\begin{equation} \label{eqn9B}
v_{pf}= \frac{3}{2} \ell L_0 \Delta g(\bar{c}_0).
\end{equation}
To ensure the phase-field model captures the physical interface kinetics, we equate the phase-field velocity $v_{pf}$ to the sharp-interface velocity $v_n$ in Eq. (\ref{eqn1B}). The expression for the phase-field mobility $L_0$ then reads
\begin{equation} \label{eqn9B}
L_0 = \frac{2}{3} \frac{v_n}{\ell\Delta g (\bar{c}_0)}= \beta i_a \quad\quad\quad \beta=  \frac{2}{3}\frac{M}{\ell \Delta g (\bar{c}_0) z F \rho},
\end{equation}
where $\beta$ is the proportionality constant that relates the phase-field mobility to the current density. In environments in which sulfate is abundant ($\bar{c}_0 \gg \bar{S}_\mathrm{min}$) and normalising the concentrations with $S_\mathrm{min}$ as in the present investigation (Section \ref{sec25}), the proportionality constant simplifies to 
\begin{equation} \label{eqn9B2}
\beta \approx \frac{2}{3}\frac{M}{\ell z F \rho} \frac{K_\mathrm{m}b}{N \Psi (Yq-b)c_0}.
\end{equation}
Using experimental data on the pitting depth of 420 $\upmu$m over 150 days (i.e., $v_n = 3.24 \times10^{-11}$ m/s, Fig. \ref{Fig3}) and computing $\Delta g(\bar{c}_0)$ from Eq. (\ref{eqn6B}) for the reference bulk concentration of $c_0$ = 0.5 mM used in the experiment (Section \ref{sec3}), returns a value of $L_0 = 1.13 \times 10^{-12}$ m$^3$/(J$\cdot$s), which aligns well with the calibrated value of $L_0 = 2.10 \times 10^{-12}$ m$^3$/(J$\cdot$s) in Section \ref{sec3}. The difference between these two values is expected since the interface has a physical thickness and sulfate concentration varies within the diffuse interface zone. This local concentration depletion reduces the actual driving forces, requiring a higher mobility parameter in the simulation. Albeit crude, this sharp interface analysis provides an order of magnitude estimation and a lower bound of the phase field mobility parameter. Relation (\ref{eqn9B}) can be applied to abiotic phase-field corrosion models with the caveat that the driving force $\Delta g$ has a different form. This analysis applies to conditions in which corrosion is driven by short-range interactions (activation-controlled corrosion).

\begin{singlespace}

\end{singlespace}
\end{document}